\documentclass[twocolumn,showpacs,aps,prb,floatfix,superscriptaddress,noshowpacs]{revtex4-1}
\usepackage[latin9]{inputenc}
\usepackage{amsmath}
\usepackage{cancel}
\usepackage{graphicx}
\usepackage{esint}
\PassOptionsToPackage{normalem}{ulem}
\usepackage{ulem}
\usepackage{verbatim}

\makeatletter
\@ifundefined{textcolor}{}
{%
 \definecolor{BLACK}{gray}{0}
 \definecolor{WHITE}{gray}{1}
 \definecolor{RED}{rgb}{1,0,0}
 \definecolor{GREEN}{rgb}{0,1,0}
 \definecolor{BLUE}{rgb}{0,0,1}
 \definecolor{CYAN}{cmyk}{1,0,0,0}
 \definecolor{MAGENTA}{cmyk}{0,1,0,0}
 \definecolor{YELLOW}{cmyk}{0,0,1,0}
}

\usepackage{amsfonts}
\usepackage{amsmath}
\usepackage{booktabs}
\usepackage{url}
\usepackage{graphicx}
\usepackage{color}
\usepackage[usenames,dvipsnames,svgnames,table]{xcolor}

\newcommand{\vect}[1]{\boldsymbol{\mathbf{#1}}}

\newcommand{\eref}[1]{Eq.~\ref{#1}}

\newcommand{\cref}[1]{Chp.~\ref{#1}}

\newcommand{\be}{\begin{equation}}
\newcommand{\ee}{\end{equation}}
\newcommand{\beq}{\begin{eqnarray}}
\newcommand{\eeq}{\end{eqnarray}}

\usepackage{color}
\usepackage{ulem}
\usepackage{dsfont}

\definecolor{violet}{rgb}{0.58, 0.0, 0.83}

\makeatother

\begin{document}

\title{Non-adiabatic bulk-surface oscillations in driven topological insulators}

\author{Michael\ \surname{Kolodrubetz}}
\affiliation{Department of Physics, University of California, Berkeley,
California, 94720, USA} 
\affiliation{Materials Sciences Division, Lawrence Berkeley National Laboratory, 
Berkeley, California 94720, USA}

\author{Benjamin\ \surname{M. Fregoso}}

\affiliation{Department of Physics, University of California, Berkeley,
California, 94720, USA}

\author{Joel\ \surname{E. Moore}}
\affiliation{Department of Physics, University of California, Berkeley,
California, 94720, USA} 
\affiliation{Materials Sciences Division, Lawrence Berkeley National Laboratory, 
Berkeley, California 94720, USA}

\begin{abstract}
Recent theoretical and experimental work has suggested the tantalizing possibility of opening a
topological gap upon driving the surface states of a three-dimensional strong topological insulator
(TI) with circularly polarized light.  With this motivation, we study the response of TIs to a driving
field that couples to states near the surface. We unexpectedly find coherent oscillations between
the surface and the bulk and trace their appearance to unavoidable resonances caused by photon
absorption from the drive. We show how these resonant oscillations may be captured by the Demkov-
Osherov model of multi-level Landau-Zener physics, leading to non-trivial consequences such as
the loss of adiabaticity upon slow ramping of the amplitude. We numerically demonstrate
that these oscillations are observable in the time-dependent Wigner distribution, which is directly
measurable in time-resolved ARPES experiments. Our results apply generically to any system with
surface states in the presence of a gapped bulk, and thus suggest experimental signatures of a novel
surface-bulk coupling mechanism that is fundamental for proposals to engineer non-trivial states by
periodic driving.
\end{abstract}

\maketitle

The recent emergence of topological physics in bulk materials has
brought to bear an important connection between topology in the bulk
and novel surface states. These surface states manifest a variety
of interesting properties, such as exhibiting anomalous behavior that
is impossible in a purely two-dimensional theory.
The simplest example of this is the one-dimensional chiral edge states
in the quantum Hall effect \cite{Klitzing1980_1,Halperin1982_1,Tsui1982_1,Wen1992_1},
and the same concept applies to helical surface states and isolated Dirac cones in
two- and three-dimensional topological insulators respectively \cite{Kane2005_1,Bernevig2006_1,Konig2007_1,Moore2007_1,Fu2007_1,Fu2007_2,Hsieh2008_1},
as well as more exotic cases like Fermi arcs in Weyl and Dirac semimetals \cite{Murakami2007_1,Wan2011_1,Liu2014_1,Xu2015_1,Xu2015_2,Lu2015_1}.
Indeed, an ever-expanding zoo of surface states is continuously being
discovered \cite{Kitaev2001_1,Moore2008_1,Levin2009_1,Kitagawa2010_1,Oreg2010_1,Lutchyn2010_1,Fu2011_1,Tanaka2012_1,Dziawa2012_1,Mourik2012_1,Okada2013_1,TitumArxiv2015_1}.

These surface states are particularly amenable to detection by a host
of modern experimental methods, such as scanning tunneling microscopy
(STM) \cite{Seo2010_1, Gyenis2013_1, Nadj-Perge2014_1, Jeon2014_1, Zeljkovic2015_1, Inoue2016_1} 
and angle-resolved photoemission spectroscopy (ARPES) \cite{Hsieh2008_1, Dziawa2012_1, Xu2015_1, Lu2015_1, Xu2015_2}. 
These probes preferentially excite electrons near the surface and are thus
able to measure and distinguish surface and bulk states. A more recent
development in ARPES as well as similar photon-in photon-out
experimental setups is time-resolved pump-probe spectroscopy, in which
the system is be excited far from equilibrium and the state detected
during the relaxation process \cite{Parker1972_1, Hsieh2011_1, Liu2011_1, Smallwood2012_1, Wang2012_1,Wang2013_1, Hu2014_1,Kaiser2014_1,Neupane2015_1}. This gives much richer insight into
both the static and dynamic properties of the quantum system and has
also given rise to a recent re-emergence of theory for such far-from-equilibrium
systems. In particular, there is an active search for examples of drive-induced topological phases
\cite{Oka2009_1,Lindner2011_1,Kitagawa2011_1}
and significant theoretical progress towards their classification\cite{Rudner2013_1,Carpentier2015_1,Roy_1,Potter_1,Keyserlingk_1,Else_1}. 

One important development in the field has been a recent experiment\cite{Wang2013_1} 
in which a time-reversal-invariant topological insulator (TI) was irradiated with a pulse of light and imaged via pump-probe ARPES.
The Dirac cone on the surface of these materials is a seed of new topological physics, and the experiment
sees a gap open in the Dirac cone upon applying circularly polarized light.
This gap is predicted to be topological in the sense that it realizes a half-integer quantum Hall effect. \cite{Qi2008_1, Essin2009_1}

Motivated by this development,
in this paper we explore the non-equilibrium dynamics of a topological insulator in the presence of a short pulsed drive. 
The pulse breaks the perfect periodicity of the drive, yet we numerically see
Floquet-Bloch sidebands as in the experiments. However, we find an unexpected oscillation
in the intensity of these sidebands, which we identify as a novel bulk-surface coupling induced
by the local drive at the surface. We show that this coupling leads to \textit{coherent} 
oscillations between the surface and the bulk that survive in the thermodynamic limit, which generically arise through a simple many-level
Landau-Zener picture that depends on Floquet resonances. This model yields several non-trivial predictions,
including reversing the meaning of adiabaticity its traditional non-resonant behavior: 
faster ramps appear more ``adiabatic'' because they see the resonances for less time, and thus decreasing
the ramp rate leads to stronger bulk-surface oscillations. We find that these resonant oscillations are 
not only visible in the Wigner distribution, a non-equilibrium observable measurable in pump-probe ARPES,
but are completely generic to periodically-driving the surface of any material with surface states
inside a gapped bulk. This provides a measurable signature of this non-trivial surface-bulk resonance
that should play a major role in Floquet engineering of driven surface states.

The paper is organized as follows. In Section~\ref{sec:FB_review} we introduce the idea of 
Floquet-Bloch states and a non-equilibrium observable -- the Wigner distribution --
that can be used to measure them. We then discuss the behavior of these states at constant amplitude of
drive for the simplest single-Dirac-cone model of TI surface states followed by a more complicated
model in which coupling is allowed to the bulk. In Section~\ref{sec:non_adiabatic_pulse} we see 
how the Floquet-Bloch states are modified by turning the drive on and off non-adiabatically via a 
Gaussian pump pulse. One important effect that we see is resonance between the surface and bulk, 
which we proceed to describe using a many-level generalization of Landau-Zener tunneling known as 
the Demkov-Osherov model. Finally, in Sec.~\ref{sec:arpes_fapt}, we analytically derive other 
leading corrections to the adiabatic Floquet-Bloch signal using the Floquet generalization of adiabatic perturbation theory.
  
\section{Floquet-Bloch states for constant amplitude drive}
\label{sec:FB_review}

\begin{figure}
\includegraphics[width=1\columnwidth]{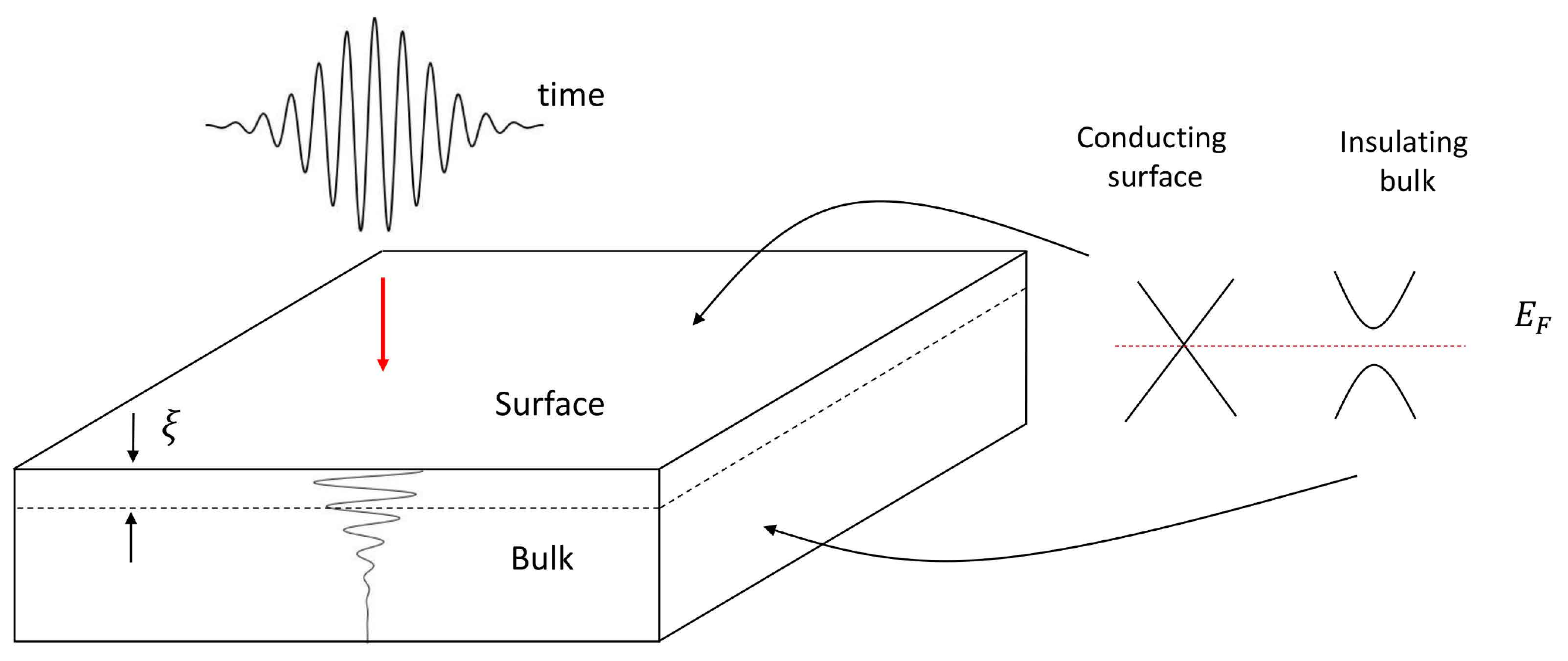}

\caption{\label{fig:setup_plus_ti_band_struct} Illustration of the setup
that we consider. A pulsed periodic electric field is incident on
the surface of a topological insulator and decays with some characteristic
length scale $\xi$ into the bulk. Also shown the band structure of a topological
insulator (TI). The bulk appears to be a gapped semiconductor, but
the difference in the topological $\mathbb{Z}_{2}$ invariant between
the TI and the surrounding vacuum leads to Dirac cone surface states.}
\end{figure}

A schematic setup used in many contemporary condensed matter experiments
is illustrated in Fig. \ref{fig:setup_plus_ti_band_struct}. A laser
pulse illuminates the sample, driving the electrons out of equilibrium.
The non-equilibrium electrons are then measured via one of a number
of methods, e.g., optical response, photoemission, tunneling, etc.
This type of setup is particularly interesting in the case of topological
insulators, whose surface states may be readily excited by the 
drive. In addition, as the bulk states have
some (weak) overlap with the drive, they also are excited. This is
precisely the effect that is used in pump-probe experiments of high-temperature
superconductors and other materials, where the non-equilibrium (bulk)
population in excited states is seen to decay as a probe of the material's
physics.

We will examine the response of topological insulators to this type
of drive. These materials have a gapped
bulk and conducting Dirac-like surface states, as illustrated in
Fig.~\ref{fig:setup_plus_ti_band_struct}. The surface states
have been probed through a number of techniques including pump-probe ARPES. 
However, due to their gapped nature, understanding the connection between 
the surface and the bulk states has remained relatively unexplored area. 
In this paper we will show that an interesting
connection exists and discuss its observable consequences.

\subsection{Driving surface states of TIs}
\label{sec:drive_surface_states}

The simplest model of a TI surface state is a single Dirac cone with
Hamiltonian \cite{Fu2007_1}
\begin{equation}
H_{\mathrm{SS}}=-v(k_{x}\sigma^{x}+k_{y}\sigma^{y}),
\label{eq:H_dirac}
\end{equation}
for a surface perpendicular to $\hat{z}$. The Pauli matrices often correspond
to physical spin $(S^x,S^y)=(\sigma^y,-\sigma^x)/2$ which is locked perpendicular to the momentum
$\vect k_\parallel$ via Rashba spin-orbit coupling \cite{Hsieh2009_1, Xu2011_1, Jozwiak2011_1}, though more generally $\sigma$ could 
denote spin/orbital indices. Our units are set by 
velocity $v=1$, as well as $\hbar=1$ throughout the paper. Consider
driving this Hamiltonian by a laser perpendicular to the surface,
with electric field $\vect E=E_{x}\cos(\Omega t+\varphi_{x})\hat{x}+E_{y}\sin(\Omega t+\varphi_{y})\hat{y}$
of constant amplitude.
This drive allows arbitrary polarization, but we will focus on the
case of linearly $y$-polarized light and phase
$\varphi_{y}=0$. Coupling this periodic drive to the surface states
is achieved by the minimal substitution $\vect k_\parallel\to\vect k_\parallel-e\vect A$,
where we pick the gauge $\vect E=-\partial \vect A/\partial t$. 
Then the Hamiltonian becomes time-dependent:
\begin{equation}
H(t)=H_\mathrm{SS} (\vect k_\parallel - e \vect A (t))~.
\end{equation}
We first consider the case of constant
drive amplitude, $E_{y}=\Omega A_{y}$,
but later we will return to the case where this amplitude in
turn varies slowly as in the case of a pulsed laser 
(cf. Fig.~\ref{fig:driven_ss_wigner_distr}b). Note that we
are assuming the drive is uniform over the entire sample such that
the in-plane momentum $\vect k_\parallel$ remains a conserved quantity even in the presence
of the drive.

\subsubsection{Non-equilibrium observables: Wigner distribution and Floquet-Bloch
states}

Acting on the Hamiltonian in Eq. \ref{eq:H_dirac} with a periodic
drive yields a fundamentally non-equilibrium problem. Floquet's theorem states that the full time
evolution $U(t)=\mathcal{T}\exp[-i\int H(t)dt]$ ($\mathcal{T}$ =
time ordering) can be decomposed as 
\begin{align}
U_{\vect k\parallel}(t)=&P_{\vect k\parallel}(t)e^{-iH_{F}(\vect k\parallel)t},
\label{eq:floquet_bloch_thm}
\end{align}
where $P_{\vect k\parallel}(0)=\mathds1$ and $P_{\vect k\parallel}(t)=P_{\vect k\parallel}(t+2\pi/\Omega)$.
$P$ is a periodic operator often called the micromotion 
and $H_{F}^{k\parallel}$ is an effective static Hamiltonian - the Floquet
Hamiltonian - that describes the behavior over many cycles. This is
the temporal analogue of Bloch's theorem, in addition to which we
have used the usual Bloch's theorem in noting that $\vect k_\parallel$ is conserved.
The eigenstates of $H_{F}^{k\parallel}$ satisfying $H_{F}(\vect k_\parallel)|n_{F}(\vect k_\parallel)\rangle=
\epsilon_{F}^{n}(\vect k_\parallel)|n_{F}(\vect k_\parallel)\rangle$
are known as Floquet-Bloch states \cite{Wang2013_1,Fregoso2013_1,Mahmood2016_1}. A system prepared in one of these
Floquet-Bloch states at time $t=0$ will return to the same state
stroboscopically at times $t=nT$ for integer $n$, where $T=2\pi/\Omega$
is the driving period.

For such a non-equilibrium system, one of the most natural observables
to consider is the probability to be in each of the Floquet-Bloch
eigenstates. This is naturally described by the
non-equilibrium generalization of the occupation number, namely the
Wigner distribution:
\begin{eqnarray*}
f_{\alpha\beta}(\vect k_\parallel,\omega,t_\mathrm{av}) & = & -iG_{\alpha\beta}^{<}(\vect k_\parallel,\omega,t_\mathrm{av})\\
 & = & -i\mathcal{F}_{t_r}\left[\langle c_{\beta}^{\dagger}(t_\mathrm{av}+t_\mathrm{r}/2)c_{\alpha}(t_\mathrm{av}-t_\mathrm{r}/2)\rangle \right],
\end{eqnarray*}
where $\alpha$ and $\beta$ denote spin/orbital indices and $\mathcal{F}$
is the Fourier transform. To ensure basis-independence
we will be interested in its trace, $f(\vect k_\parallel,\omega,t_\mathrm{av})\equiv\sum_{\alpha}f_{\alpha\alpha}(\vect k_\parallel,\omega,t_\mathrm{av})$.

The Wigner distribution naturally describes equilibrium or non-equilibrium
occupation of energy eigenstates\cite{Kadanoff1989_1}. The simplest example of this is
to consider evolution of the state $|\psi\rangle=c_{0}^{\dagger}|vac\rangle$
under a static single-particle Hamiltonian, where $c_{0}^{\dagger}$
creates a single fermion in the energy eigenvalue $E_{0}$ of $H$.
Then a straightforward calculation confirms that $f$ is just a single
peak at frequency $E_{0}$:

\begin{eqnarray}
f(t_\mathrm{r},t_\mathrm{av}) & = & \langle\psi|\sum_{n}c_{n}^{\dagger}(t_\mathrm{av}+t_\mathrm{r}/2)c_{n}(t_\mathrm{av}-t_\mathrm{r}/2)|\psi\rangle\nonumber \\
 & = & \sum_{n}\langle\psi|\Big(e^{iH(t_\mathrm{av}+t_\mathrm{r}/2)}c_{n}^{\dagger}e^{-iH(t_\mathrm{av}+t_\mathrm{r}/2)}\nonumber \\
 &  & \;\;\;\;\;\;\;\;e^{iH(t_\mathrm{av}-t_\mathrm{r}/2)}c_{n}e^{-iH(t_\mathrm{av}-t_\mathrm{r}/2)}\Big)|\psi\rangle\nonumber \\
 & = & e^{iE_{\psi}t_\mathrm{r}}\sum_{n}\langle\psi|c_{n}^{\dagger}e^{-iHt_\mathrm{r}}c_{n}|\psi\rangle=e^{iE_{0}t_\mathrm{r}}\nonumber \\
f(\omega,t_\mathrm{av}) & = & 2\pi\delta(\omega-E_{0}).\label{eq:wigner_distr_energy_eig}
\end{eqnarray}
Similarly, if we start with many electrons, $|\psi\rangle=c_{0}^{\dagger}\cdots c_{N-1}^{\dagger}|vac\rangle$,
then a similar calculation shows that $f$ is just a sum
of peaks at each electron's energy: $f(\omega)=2\pi\sum_{j=0}^{N-1}\delta(\omega-E_{j})$.
Thus the Wigner distribution gives information about not only the
occupation via the amplitude of the delta-function peaks ($2\pi$
per electron), but also about their time-evolution via the peak frequency.

These ideas are particularly useful in driven Floquet systems as they
are out-of-equilibrium from the get go. Before deriving the Wigner
distribution of a system in a Floquet eigenstate, let's start by
considering the more generic case where one starts in an eigenstate
of some static $H$ at time $t_{0}$
but then turns on an arbitrary driving $H(t)$. As long as the Hamiltonian
remains non-interacting, by Wick's theorem the Wigner distribution
will remain the sum over occupied eigenstates of the single-particle
$f$. So if we start from some single-particle state $|\psi_{n}(t_{0})\rangle\equiv c_{n}^{\dagger}|vac\rangle$
and then turn on arbitrary drive, it is readily confirmed that $f$ is simply
given by
\begin{equation}
f_{n}(t_\mathrm{r},t_\mathrm{av})=\langle\psi_{n}(t_\mathrm{av}+t_\mathrm{r}/2)|\psi_{n}(t_\mathrm{av}-t_\mathrm{r}/2)\rangle,\label{eq:wigner_fn_overlap}
\end{equation}
where $|\psi_{n}(t)\rangle=U(t,t_{0})|\psi_{n}(t_{0})\rangle$ is
the state obtained by full time evolution starting from $|\psi_{n}(t_{0})\rangle$.
For $N$ occupied single particle states one simply sums over $n=0,1,\ldots,N-1$.

Now consider a Floquet-Bloch eigenstate $|n_{F}(\vect k_\parallel)\rangle$. As we work
with translationally-invariant drives throughout this paper, we will
occasionally suppress the $\vect k_\parallel$ dependence.
Associated with a given Floquet eigenstate are a time-periodic family
of wave functions,
\begin{equation}
|n_{F}(t)\rangle\equiv P(t)|n_{F}\rangle,\label{eq:floquet_eig_time_dep}
\end{equation}
which describe how $|n_{F}\rangle$ evolves during a cycle. Note that
by our convention for $P$, $|n_{F}(0)\rangle=|n_{F}\rangle$. As
this state is periodic, we may Fourier decompose it:
\begin{equation}
|n_{F}(t)\rangle=\sum_{\ell}e^{i\ell\Omega t}|n_{F}^{(\ell)}\rangle.
\label{eq:n_F_ell}
\end{equation}
These Floquet modes $|n_{F}^{(\ell)}\rangle$ play an important role
in the theory. In particular, if we plug the Floquet eigenstate into
Eq. \ref{eq:wigner_fn_overlap}, we see that 
\begin{eqnarray*}
f_{n}(t_\mathrm{r},t_\mathrm{av}) & = & \langle\psi_{n}(t_\mathrm{av}+t_\mathrm{r}/2)|\psi_{n}(t_\mathrm{av}-t_\mathrm{r}/2)\rangle\\
 & = & \sum_{\ell\ell^\prime}e^{i(\ell-\ell^\prime)\Omega t_\mathrm{av}}e^{i[\epsilon_{F}^{n}-(\ell+\ell^\prime)\Omega/2]t_\mathrm{r}}\langle n_{F}^{(\ell^\prime)}|n_{F}^{(\ell)}\rangle,
\end{eqnarray*}
where $|\psi_{n}(t)\rangle=e^{-i\epsilon_{F}^{n}t}|n_{F}(t)\rangle$
accounts for time evolution due to both micromotion and the Floquet
Hamiltonian (see Eq. \ref{eq:floquet_bloch_thm}). This expression
simplifies even further in an important limit, namely when we average
over the ``measurement time'' $t_\mathrm{av}$. This naturally emerges
in a number of physically-relevant situations. For instance, if we
put back in the phase of the drive, which enters the previous expression
as $\Omega t_\mathrm{av}\to\Omega t_\mathrm{av}+\varphi$, then averaging over the
often experimentally-uncontrolled phase is equivalent to averaging
over $t_\mathrm{av}$. Equivalently, one often finds that there is experimental
imprecision on the time of measurement and/or the relative phase of
the pump and the probe. If this imprecision is long
compared to the drive period, again the averaging emerges. Denoting
this so-called Floquet non-stroboscopic (FNS \cite{Bukov2014_1})
averaging by an overline, we see that \cite{Dehghani2014_1,FarrellArxiv2016_1}
\begin{eqnarray*}
\overline{f_{n}(t_\mathrm{r})} & = & \sum_{\ell}e^{i(\epsilon_{F}^{n}-\ell\Omega)t_\mathrm{r}}\langle n_{F}^{(\ell)}|n_{F}^{(\ell)}\rangle\\
\overline{f_{n}(\omega)} & = & 2\pi\sum_\ell \delta(\omega-\epsilon_{F}^{n}+\ell\Omega)\langle n_{F}^{(\ell)}|n_{F}^{(\ell)}\rangle.
\label{eq:f_n_omega_bar}
\end{eqnarray*}
So each electron state is ``split'' into Fourier modes at frequency $\epsilon_{F}^{n}-\ell\Omega$
with amplitude $p_{n\ell}=\langle n_{F}^{(\ell)}|n_{F}^{(\ell)}\rangle$.
Note that these peaks sum up to 1 total electron, $\sum_{\ell}p_{n\ell}=1$,
by the normalization of $|n_{F}\rangle$. So we see that the Wigner
distribution again provides insight on the frequency of these sidebands
as well as the probability to occupy them.

\begin{figure*}
\includegraphics[width=.8\textwidth]{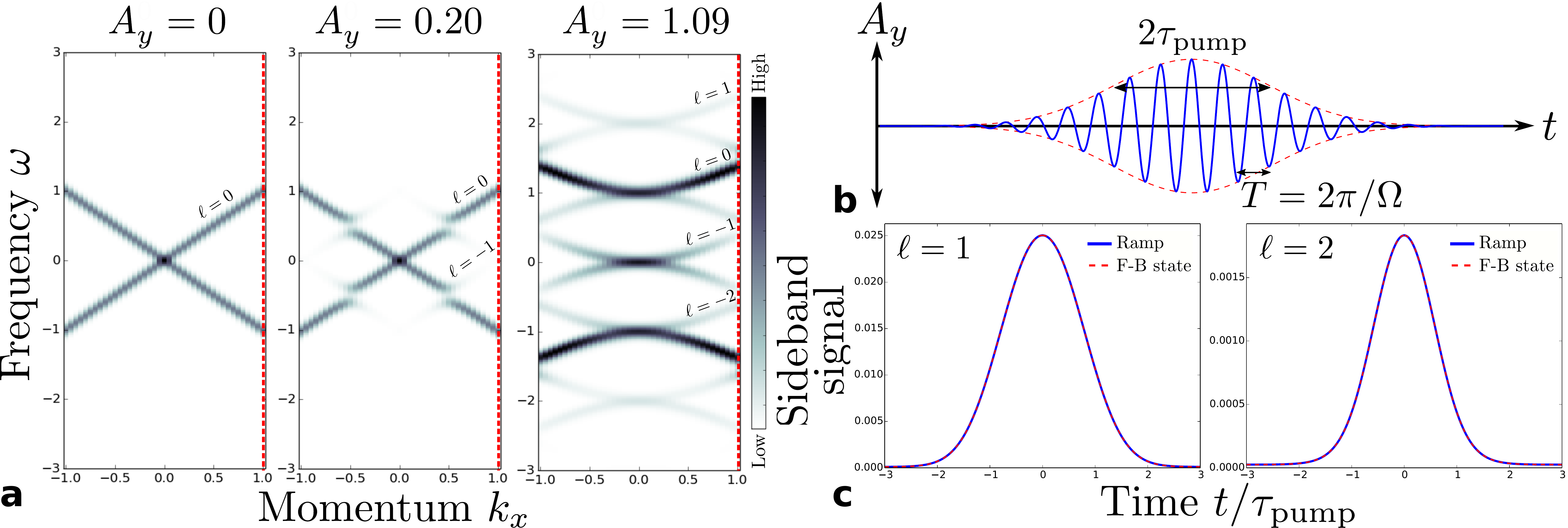}
\caption{\label{fig:driven_ss_wigner_distr}Driving Dirac surface
states of a TI model (Eq.~\ref{eq:H_dirac}) without any bulk states. 
(a) Wigner distribution of ``equilibrium'' Floquet-Bloch
states with varying drive strength $A_y$, as derived in Eq.~\ref{eq:f_n_omega_bar},
showing hybridization of the surface states.
The chemical potential $\mu > 0$ is chosen such that 
both surface states are occupied.  Delta-function peaks corresponding to Floquet-Bloch states have been
broadened by a Gaussian of width $\tau_\mathrm{pr}=24$ for clarity.
(b) Illustration of the pulsed drive that we will consider.
For the driven Dirac model, both states in the Hilbert space are occupied, so in the slow-ramp
limit ($\tau_\mathrm{pump} \gg T$) there are no additional excitations due to the ramp. This
is illustrated in (c) by plotting the amplitude in the $\ell=1,2$ sidebands for
$k_x=1$, which show no difference
between the instantaneous Floquet-Bloch eigenstates (dashed red) and the full time evolution 
during the ramp (blue). Data is for a linearly-polarized pump perpendicular to the momentum, with
parameters $\tau_\mathrm{pump}=60$, $A_y^0=1.09$, and $\Omega=1$.}
\end{figure*}

Let us now apply these ideas to driving the surface states of the
TI, described by the Hamiltonian in Eq. \ref{eq:H_dirac}. As we have
shown, the signal at each $\vect k_\parallel$ is just the sum over the signals
from each of the occupied states. In Fig. \ref{fig:driven_ss_wigner_distr}a
we plot the Wigner distribution in the Floquet eigenstates 
with both branches of the Dirac cone occupied for distinct (but constant in time) 
drive amplitude. For the remainder of the paper, we focus on 
linearly-polarized light whose polarization direction ($\hat y$) is orthogonal
to the momentum ($k_y=0$). Other choices of polarization and momentum give qualitatively
similar results. As noted in the plots and seen 
elsewhere\cite{Oka2009_1,Syzranov2008_1,Kitagawa2011_1,Fregoso2013_1,Sentef2015_1}, anti-crossings between
the surface states occur open up near the resonance between the branches.
This is the first example we will see of Floquet resonance, here between
two surface states. These Floquet resonances, and in particular more
complicated ones between the surface and the bulk, will play a starring
role in the remainder of the paper.

As we will discuss in more detail later, actual experiments involve a pulsed rather
than fixed drive, as illustrated in Fig. \ref{fig:driven_ss_wigner_distr}b. In the 
slow ramp limit, $\tau_\mathrm{pump} \gg T$, which we always restrict ourselves to,
the drive is approximately periodic at any point in time and we might expect the system
to adiabatically follow the instantaneous Floquet-Bloch eigenstates. In general, if we ramp too fast, we expect
non-adiabatic effects as we fail to adiabatically
follow these eigenstates. However, we note that because both bands are occupied,
there are \emph{no} non-adiabatic effects in this purely surface state model no matter how short the
pulse. The reason is simply that both states in the two-level system
are filled, and there is simply nowhere else in the Hilbert states
for the electrons to go. This is seen in Fig.~\ref{fig:driven_ss_wigner_distr}c, where
the $\ell$th sidebands of the Wigner distribution of the Floquet-Bloch eigenstates are 
compared to those of the full time evolution, showing no difference for $\tau_\mathrm{pump} \gg T$.

We are primarily interested in non-adiabatic effects in the periodically-driven system 
due to the pulse. We will show that such novel behavior can occur
when coupling these states to an empty bulk conduction band.
Therefore, let us now consider the presence
of the bulk and see how it affects this story.

\subsection{Driving surface and bulk states of TIs}

To understand the relevance of the bulk, we want to start by constructing
a simple tight-binding model of a three-dimensional topological insulator.
We consider one of the simplest such bulk models \cite{Qi2008_1}, namely the lattice
regularization of $(\vect k\cdot\vect\sigma)\tau^{z}+m\tau^{x}$:
\begin{eqnarray}
H_{\mathrm{bulk~ TI}} & = & (\sigma^{x}\sin k_{x}+\sigma^{y}\sin k_{y}+\sigma^{z}\sin k_{z})\tau^{z}+\nonumber \\
 &  & (m+3-\cos k_{x}-\cos k_{y}-\cos k_{z})\tau^{x},\label{eq:H_bulk_ti}
\end{eqnarray}
where $\sigma$ and $\tau$ are two sets of Pauli matrices corresponding
to, e.g., spin and orbital degrees of freedom. We again assume that
the electric field couples via the minimal substitution, but now with
the caveat that the electric field strength decays into the bulk with
length scale $\xi$, as in Fig.~\ref{fig:setup_plus_ti_band_struct}. Choosing the surface of interest to again be
perpendicular to $\hat{z}$, $\vect k_\parallel=(k_x,k_y)$
remain good quantum numbers. For more details of the hopping Hamiltonian in the $z$-direction, please see
Appendix~\ref{sec:bulk_ti_model_details}.

In the absence of drive, this model gives a topological insulator
for $-4<m<0$ and a trivial insulator otherwise. 
In the presence of drive, we can solve this Floquet problem and calculate
its Wigner distribution. The results are shown in Fig.~\ref{fig:driven_ss_plus_bulk}a.
Similar to the simple surface-only model, the surface states are strongly
dressed by the drive, although details of the signal depend heavily on
microscopic details of the model. At strong driving strength $A_y=0.25$, 
this ``Floquet equilibrium'' (i.e., constant drive
amplitude) data already shows how coupling to the bulk changes the story,
resulting in $\ell=3$ and $4$ sidebands that are stronger than $\ell=2$ due
to resonant surface-bulk hybridization. In the next section we will
see that this surface-bulk coupling has a strong effect when we consider a pulsed drive.

\section{Non-adiabatic effects of the pulsed drive}
\label{sec:non_adiabatic_pulse}

While the previous section considered the response in Floquet eigenstates,
which would come for example from the steady-state of a continuous-wave
laser, it is often more experimentally practical to use 
pulsed sources. This has undesirable effects such as losing perfect
periodicity, but the ability to change the pulse length can also be a powerful tool to prevent 
heating and target a unitary response of the system. Therefore, in this section we will concern ourselves with
the question of how finite pulse width affects the non-equilibrium
observables of driven topological insulators.

\begin{figure*}
\includegraphics[width=.8\textwidth]{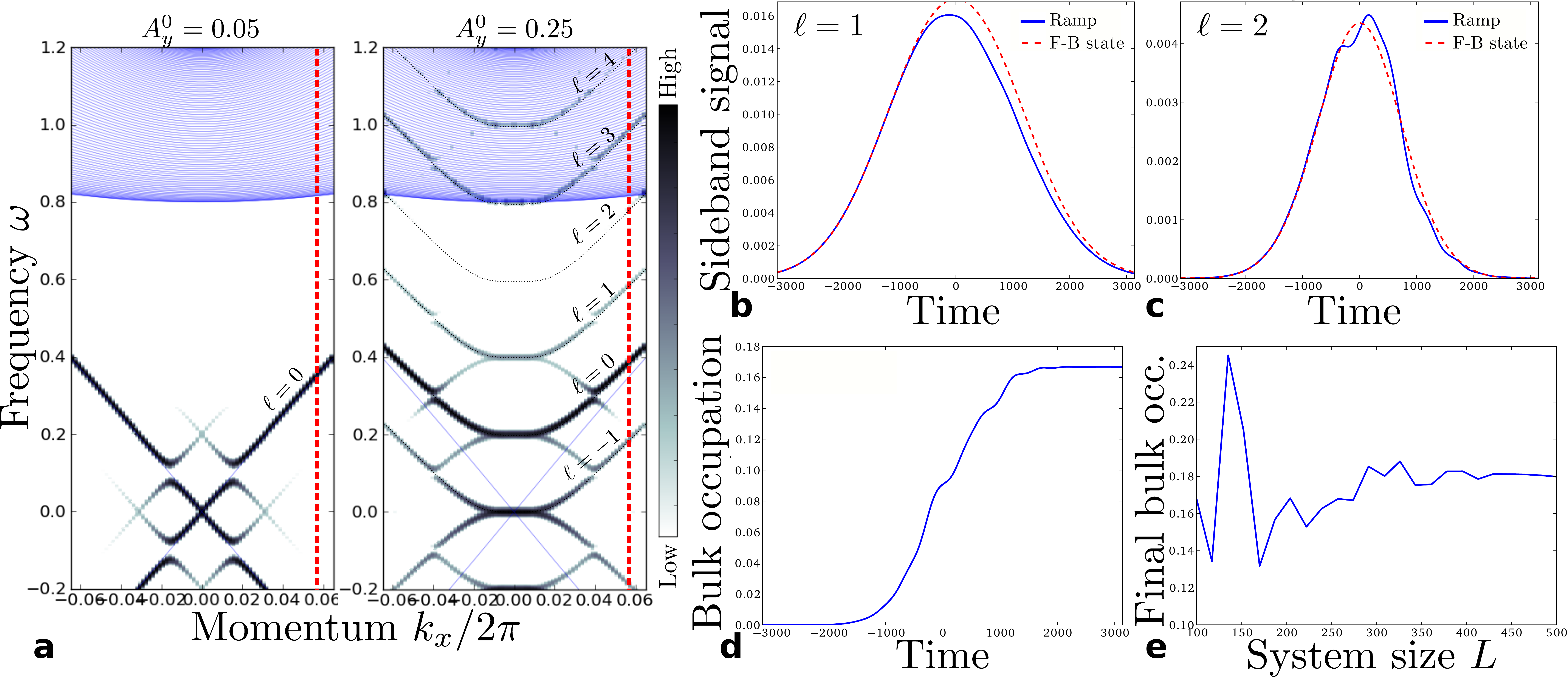}
\caption{\label{fig:driven_ss_plus_bulk}Driving surface
plus bulk states of a model TI (see Eq. \ref{eq:H_bulk_ti} and following text). 
(a) Wigner distribution of Floquet-Bloch eigenstates as a function of driving
amplitude. The undriven energy spectrum is shown in blue. The surface states
appear similar to Fig.~\ref{fig:driven_ss_wigner_distr}a, though in this figure
the color plot is on a logarithmic scale to make higher harmonics visible. We also
see that starting at approximately $\ell=3$, the surface-state harmonics appear in the bulk.
(b,c) Signal in the $\ell=1,2$ surface sidebands for $k_x/2\pi=0.056$, indicated by the red line in (a).
Unlike the driven Dirac model, there is a noticeable difference between the Floquet-Bloch
states (dashed red) and the exact time evolution (blue). The intensity in the $\ell$th harmonic at time $t_\mathrm{av}$ is given
by integrating the signal $I(\omega,t_\mathrm{av})$ from $\omega=\epsilon_{F}(\lambda(t_\mathrm{av}))+(\ell-1/2)\Omega$
to $\epsilon_{F}(\lambda(t_\mathrm{av}))+(\ell+1/2)\Omega$ while the ``equilibrium'' value 
is estimated by manually removing resonances\cite{Footnote_ARPES_remove_res}.
(d) Signal in the bulk bands as a function of $t_\mathrm{av}$, given by integrating the signal for all $\omega>E_{bulk}$.
(e) Final probability to occupy the bulk states after the ramp is finished ($t \to \infty$)
as a function of system size showing the existence of a well-defined thermodynamic limit.
All data are for $k_x/2\pi=0.056$, $k_y=0$, $A_x=0$, $A_y=0.25$, $m=-0.8$, $\Omega=0.2$, 
$\tau_\mathrm{pr}=5T=10\pi/\Omega$, and $L=100$ unless otherwise specified.}
\end{figure*}

A pulsed laser may be modeled by simply multiplying the periodic drive
by a slow envelope, such as a Gaussian: 
$\vect A(t)=[A_{x}^{0}\cos(\Omega t)\hat{x}+A_{y}^{0}\sin(\Omega t)\hat{y}]\exp(-t^{2}/2\tau_\mathrm{pump}^{2})$.
\footnote{More accurately, the electric field has this Gaussian envelope, but
we can approximate this as just a Gaussian on $\vect A$ in the limit of a long pulse,
$\tau_\mathrm{pump}\gg T$, which we consider throughout this paper.}
The envelope breaks periodicity and thus renders this no longer
an exact Floquet problem, though in the limit of a long pulse
it is approximately periodic at any given point in time. One might
then expect that the system will adiabatically track the Floquet eigenstates,
yielding Wigner distributions similar to Figs.~\ref{fig:driven_ss_wigner_distr}a
and \ref{fig:driven_ss_plus_bulk}a. This is almost
correct but, as we will now see, only part of the story. 

\subsection{Coherent bulk-surface oscillations}
\label{coherent_bulk_surface_osc}

We now simulate the coupled bulk-surface model of a TI under such
pumped drive. We start deep in the past with the drive turned off
and the chemical potential set such that all bulk valence band and
surface states are occupied. \footnote{In practice, we actually just compute time
evolution of the surface states on the upper surface as the remaining states do not affect the
signal. We have verified that occupying the valence bands does not change 
our results.} Then the exact dynamics are simulated
and the Wigner distribution $f(\omega,t_\mathrm{av})$ computed. This function is strongly peaked in $\omega$ and highly oscillatory
in $t_\mathrm{av}$ so we smooth out the results by convolving $f$ by a
Gaussian of width $\tau_\mathrm{pr}$ in both the frequency and time direction:
\begin{equation}
I(\omega,t_\mathrm{av})=\int d\omega^{\prime}dt_\mathrm{av}^{\prime}e^{-(\omega-\omega^{\prime})^{2}\tau_\mathrm{pr}^{2}}e^{-(t_\mathrm{av}-t_\mathrm{av}^\prime)^{2}/\tau_\mathrm{pr}^{2}}f(\omega^{\prime},t_\mathrm{av}^{\prime}).\label{eq:def_I_omega_tav}
\end{equation}
We refer to the result as the signal and/or intensity at frequency $\omega$
and time $t_\mathrm{av}$, which will be justified in Sec. \ref{sec:arpes}
by showing its connection to ARPES.
If the ``probe width'' $\tau_\mathrm{pr}$ is much greater than the drive
frequency, this convolution has the additional advantage of averaging
over the drive phase, such that $f$ may be replaced by $\overline{f}$
in Eq. \ref{eq:def_I_omega_tav}.

One striking difference between the equilibrium and non-equilibrium
case is that, even after the drive has been turned off, population
remains in the bulk conduction states, as seen in Fig.~ \ref{fig:driven_ss_plus_bulk}d.
This phenomenon is specific to the coupled bulk/surface model, and we do not
see it in the simpler Dirac cone model of Sec.~\ref{sec:drive_surface_states}. Decay of excited surface states
into the bulk has been anticipated in the presence of 
phonons \cite{Wang2012_1}, but note that this decay mechanism
does not exist in our model. Therefore, the population transferred
to the bulk may only come from coherent non-adiabatic processes.

In addition to tunneling into the bulk, we see coherent oscillations
in the Wigner distribution of the surface states. This is shown in
Fig.~\ref{fig:driven_ss_plus_bulk}b and c, where the signal in the
$\ell$th sideband is given by weight in the $\ell$th peak at fixed $k_x$ and $t_\mathrm{av}$
normalized by the sum over all peaks. Together, these 
results suggest that we are seeing coherent oscillations
of the population between the surface and the bulk states. 
We have varied the microscopic parameters over a wide range of values and
found that the existence of
these oscillations are remarkably robust, always appearing in tandem with an irreversible
``leaking'' into the bulk. We now seek to understand this in terms
the physics of Floquet resonances.

\subsection{Floquet resonances and Landau-Zener physics \label{sec:floquet_res_lz}}

Resonances have long been known to play a major role in Floquet systems
\cite{Weyl1916_1,Howland1989_1,Howland1989_2,Hone1997_1}. Mathematically, they come from the
fact that the drive introduces a new energy scale $\Omega$ such that
energies are only defined modulo $\Omega$. For a many-body system
of linear size $L$ in $d$ dimensions, the bare spectrum is extensive,
scaling as $L^{d}$. However, Hone et al. \cite{Hone1997_1} argued that folding by
$\Omega$ in the thermodynamic limit leads to a denser and denser
set of quasi-energy levels as the system size is increased. This in
turn leads to a dense set of weakly-avoided crossings such that even
simple ideas like tracking a single quasi-energy level to achieve
an adiabatic limit becomes ill-defined. Thus the weakly-avoided crossings,
which we call Floquet resonances, lead to a fundamental absence of
adiabaticity in Floquet systems. Furthermore, they have been suggested
to lead to heating effects \cite{Bukov2016_1} and the breakdown of high-frequency
expansions \cite{WeinbergArxiv2016_1}, which are two of the most important and active
topics in the field of Floquet engineering.

As seen in Fig.~\ref{fig:driven_ss_plus_bulk}a, Floquet
resonances between the surface and the bulk states inevitably occur in
systems such as ours, where the driving frequency $\Omega$ is less
than the bandwidth. However, there are a number of subtleties that we must consider in
comparing this to the Hone et al. result. Most notably, they were
considering coupling between bulk states due to the drive, whereas
here we are interested in coupling between the bulk and the surface
state. Since the drive primarily couples to the surface states and
only weakly to the bulk, one naive guess would be that the matrix
elements between these states would scale as the spatial overlaps
between them, $\xi/L$, vanishing in the thermodynamic limit. This
indeed seems to be the case, but one must counterbalance it against
the fact that the (one-dimensional) density of states at fixed $\vect k_\parallel$
scales as $L$. Thus these two effects conspire to create an order-1
gap in the quasi-energy spectrum which depends sensitively on various
microscopic properties. Therefore, we expect that the strength
will differ significantly from
model to model, e.g., between our simple model TI and a real material.
Nevertheless, the existence of order 1 Floquet resonances should be
robust by the above argument, and thus the phenomena we describe are completely generic.

\begin{figure}
\includegraphics[width=\columnwidth]{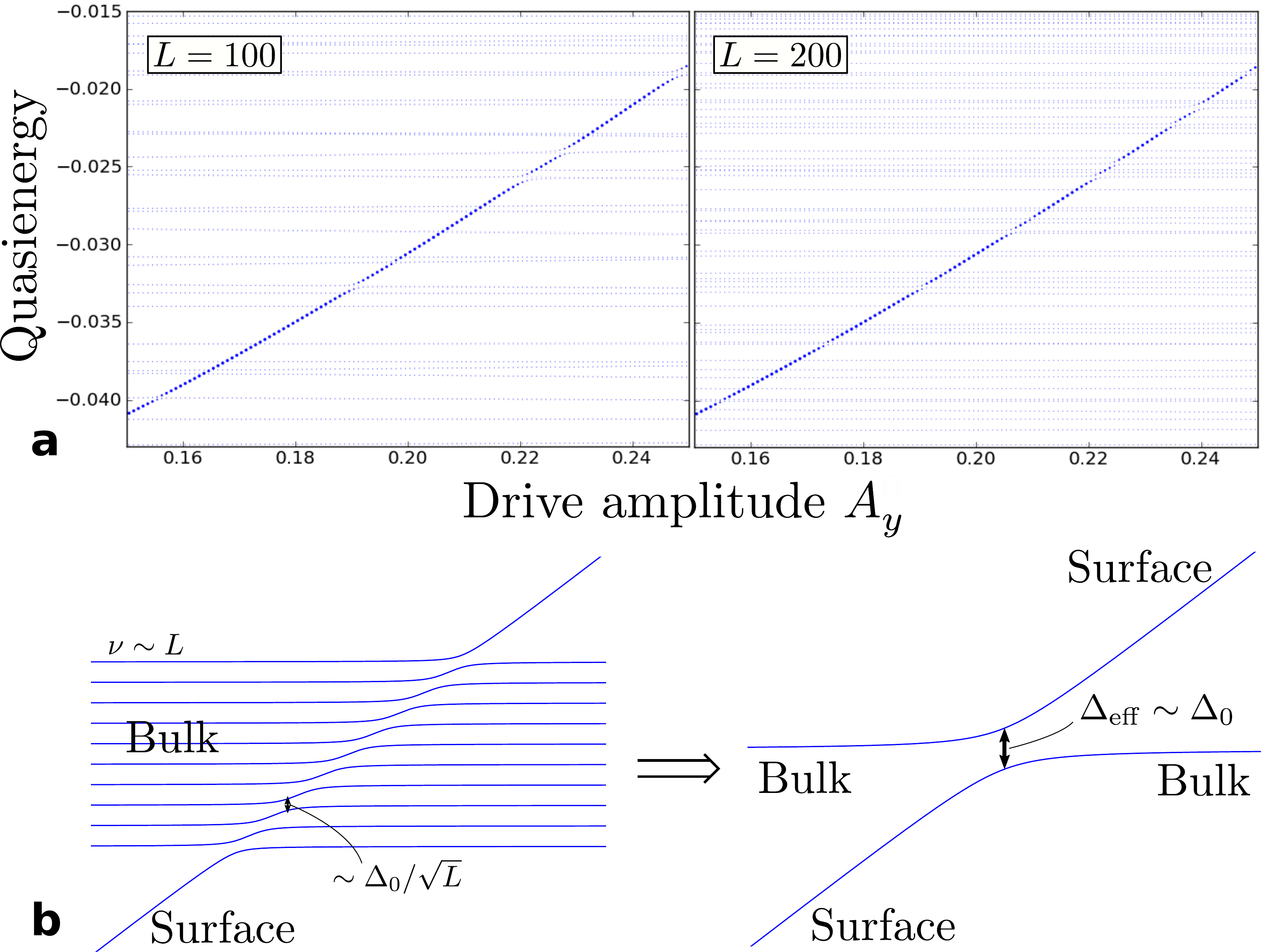}
\caption{\label{fig:illustration_resonance_lz} Floquet resonances
between the surface and bulk. (a) Floquet eigenspectrum of the TI model
as a function of drive strength using the same parameters as Fig.~\ref{fig:driven_ss_plus_bulk}b
with $L=100$ and $200$. The size of the dots is proportional to the proximity of
the Floquet eigenstate to the top surface such that the surface state appears larger
than the bulk states. (b) Illustration of the Demkov-Osherov model of the
Floquet eigenspectrum. A single surface state passes through a continuum of 
bulk states. As $L$ increases, the increase in the density of states is offset by
the decrease in the matrix elements coupling surface and bulk. The system may
be approximated by a single Landau-Zener crossing with effective gap $\Delta_\mathrm{eff}$ 
that controls both excitation of the bulk and oscillation frequency of the surface state.
See text for details.}
\end{figure}

As seen in Fig. \ref{fig:illustration_resonance_lz}a, Floquet resonances
lead to a series of anti-crossings between quasienergies of the bulk and surface states.
As expected, the quasienergy of the surface state depends strongly
on driving amplitude, while the bulk states are nearly independent
of the drive. We also confirm that as $L$ increases the number
of anti-crossings increases as well, while the strength (i.e., the gap)
of the anti-crossings decreases. This situation, where a single dispersing level passes
through many parallel non-dispersing ones is known in the non-Floquet case
as the Demkov-Osherov (D-O) model \cite{Demkov1968_1,Macek1998_1,Sinitsyn2002_1}, and is an analytically tractable many-level
generalization of the Landau-Zener (L-Z) model \cite{Landau1932_1,Zener1932_1}. The 
scattering matrix of the D-O model in the long-time limit
is remarkable because interference between the various avoided crossings is absent.
Thus, the D-O scattering problem reduces to $N_c$ independent L-Z transitions,
where $N_c$ is the number of bulk levels that the dispersing level surface
state crosses. In our case, $N_c \sim L$ at fixed
$\vect k_\parallel$ because we effectively have a one-dimensional problem.

\begin{figure*}
\includegraphics[width=.8\textwidth]{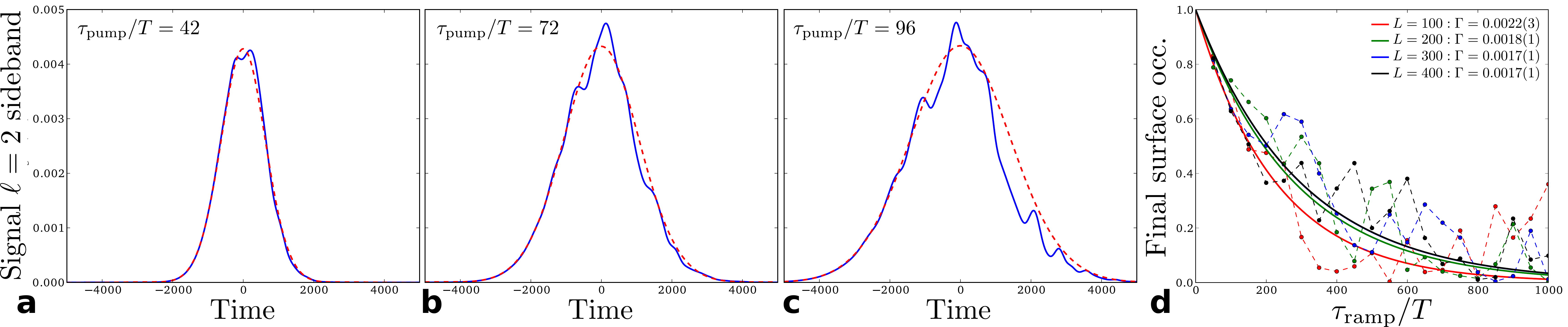}
\caption{\label{fig:signal_surface_bulk_kcut_change_params}
Scaling of surface oscillations and bulk occupation with pump time $\tau_\mathrm{pump}$, confirming
predictions of the Demkov-Osherov model. (a-c) Oscillations of the signal in the $\ell=2$ surface sideband
as a function of $\tau_\mathrm{pump}$, showing an increase in amplitude but no apparent change in the characteristic
frequency. (d) Final occupation in the surface state, which decreases exponentially with $\tau_\mathrm{pump}$
as the electrons resonantly tunnel into the bulk (see details). The decay rate into the bulk does not
depend on system size. All data are for the same parameters as Fig.~\ref{fig:driven_ss_plus_bulk}b.
}
\end{figure*}

For slow ramps, one expects that the dynamics of a Floquet system will be
dominated by resonant effects, which are captured within the appropriately-folded
effective Hamiltonian $H_F$.
Therefore, we should be able to able to treat the Floquet D-O model identically
to the undriven case. Assume the surface state is ramped through $N_c$ bulk states
during the first half of the pump pulse by increasing $A_y$ from $0$ to $A_y^0$
such that the surface state quasienergy increases at a constant velocity $v=d\epsilon/dt$. 
As each crossing may be treated independently, the final probability to be in the surface
state is just the product of the individual probabilities:
\begin{equation}
p_{ss}^A= \exp\left[-2\pi \sum_{j=1}^{N_c} \Delta_j^2 / v \right]~.
\label{eq:p_ss_A}
\end{equation}
This looks exactly like
the L-Z problem for a single avoided crossing with matrix element $\Delta_\mathrm{eff}=\sqrt{\sum_j \Delta_j^2}$,
as illustrated in Fig.~\ref{fig:illustration_resonance_lz}b. In the thermodynamic
limit, we expect these gaps to scale as $\Delta_j \sim \Delta_0 / \sqrt L$ 
from the scaling of the overlap of bulk and surface eigenstates. Thus the dynamics of our model is expected to
have a consistent $L\to\infty$ limit, which is confirmed numerically in Fig.~\ref{fig:driven_ss_plus_bulk}e.
In addition to the final bulk occupation, this effective gap also controls the time
scale of the oscillations in the surface state sidebands. Thus we see that
both the incoherent transition to bulk states and coherent bulk-surface oscillations 
survive in the thermodynamic limit with dynamics set by the same emergent energy scale.

In addition to giving a physical picture for both the surface-bulk oscillations and the
non-adiabatic tunneling of electrons into the bulk, the Demkov-Osherov model provides a handle
for understanding how these should change with the various parameters, such
as the experimentally-controllable $\tau_\mathrm{pump}$. 
One important upshot is the meaning of ``adiabaticity'' reversed from what we expect in the
absence of resonances.
Normally one expects the adiabatic limit to correspond to slow ramping, such
that the system tracks the instantaneous Floquet eigenstate. However, it is clear for the resonant case
that ramping the field too slowly will cause the entire population to transfer
into the bulk. Therefore,
to ``adiabatically'' track the surface state, one must instead use a fast ramp, though still sufficiently
slow to prevent direct non-resonant excitations to the bulk.\cite{Hone1997_1,Drese1999_1,WeinbergArxiv2016_1}
More explicitly, we expect that the population remaining in the surface state at the end of the ramp 
should scale as $p_{ss}^f = e^{-4\pi\Delta_\mathrm{eff}^2/v}\sim 
e^{-(4\pi\Delta_\mathrm{eff}^2/\Delta \epsilon)\tau_\mathrm{pump}}\equiv e^{-2\Gamma \tau_\mathrm{pump}}$, 
where the additional factor of two compared to Eq.~\ref{eq:p_ss_A} comes from ramping up to $A_y^0$ then back down to $0$.
This dependence is consistent with the data, as shown in Fig.~\ref{fig:signal_surface_bulk_kcut_change_params}d.  By a similar token,
increasing $\tau_\mathrm{pump}$ increases the size of the resonant bulk-surface oscillations, as seen in
Fig.~\ref{fig:signal_surface_bulk_kcut_change_params}a-c. 
It is interesting to note that a similar ``ghost'' surface state has been found in static models 
of topological materials coupled to a trivial bulk~\cite{Bergman2010,Baum2015}, which may be solved by
modeling it with the well-known Fano model~\cite{Fano1961}. Ours is the natural Floquet
generalization of these ideas, leading to fundamentally non-equilibrium phenomena
such as coherent bulk-surface oscillations and Floquet resonances.
Further discussion of the Demkov-Osherov model and its
application to surface-driven systems may be found in Appendix~\ref{sec:demkov_osherov_details}.

\subsection{Applications to time-resolved ARPES\label{sec:arpes}}

Before concluding this section, we note that our results our directly applicable to time-resolved ARPES experiments.
Time-resolved pump-probe ARPES works by driving the system
at frequency $\Omega$ with a Gaussian envelope (the pump) which excites that
electrons in the sample but does not cause it to photoemit. Then,
at variable times during the pump, a weak probe pulse at much higher
frequency and much shorter width $\tau_\mathrm{pr}$ is shone on the sample,
which excites the driven electrons above the work function of the
material. These electrons are then (photo)emitted by the sample and
subsequently detected. By measuring the energy and momenta of the
photoemitted electrons, the detector is able to map out the material's
band structure during the probe pulse, including any non-equilibrium
effects given by the pump.

Theoretically, the time-resolved ARPES signal for an arbitrary driven
Hamiltonian $H(t)$ is given by \cite{Freericks2009_1,Sentef2015_1}
\begin{eqnarray}
I(\omega,t_\mathrm{av}) & = & \mathrm{Im}\Bigg[\int dt_{1}dt_{2}s_\mathrm{pr}(t_\mathrm{av}-t_{1})s_\mathrm{pr}(t_\mathrm{av}-t_{2})\nonumber \\
 &  & \;\;\;\;e^{i\omega(t_{1}-t_{2})}\mathrm{Tr}G^{<}\left(t_{1}-t_{2},\frac{t_{1}+t_{2}}{2}\right)\Bigg]\label{eq:I_omega_tav_arpes}
\end{eqnarray}
if one ignores that matrix elements between the electrons in the material
and the photoemitted states. A brief discussion of the effect of non-trivial matrix
elements is found in Appendix~\ref{sec:matrix_elements_arpes}.
If one uses a Gaussian probe, $s_\mathrm{pr}(t)=\exp(-t^{2}/2\tau_\mathrm{pr}^{2})$,
then Eq. \ref{eq:I_omega_tav_arpes} reduces to Eq. \ref{eq:def_I_omega_tav}.
Thus, all of the results we have shown so far can be simply interpreted
as the signal of a time-resolved ARPES experiment with a Gaussian
pump and probe, and our results serve as an important experimentally-accessible 
signature of this novel bulk-surface coupling.

\section{Leading corrections in Floquet adiabatic perturbation theory}
\label{sec:arpes_fapt}

We have seen that, for slow pulses, non-adiabatic corrections to the Wigner distribution
are dominated by resonances between the bulk and the surface. In this section, we will
consider the other potential source of excitations, namely direct excitations
to the bulk due to fast ramping of the drive.
We will theoretically describe the leading corrections using 
Floquet adiabatic perturbation theory~\cite{Drese1999_1}, placing the Wigner distribution 
on the same footing as static observables 
(cf. \cite{WeinbergArxiv2016_1}). We will also use this to understand the
\emph{short} pump pulse limit, which remains relatively unexplored experimentally.

\subsection{Basics of Floquet adiabatic perturbation theory (FAPT)}

Adiabatic perturbation theory (APT) is a technique to derive leading
corrections to the adiabatic limit for a system with a parameter $\lambda$
that is ramped slowly with time \cite{Born1928_1,Kato1950_1,Teufel2003_1,Rigolin2008_1,DeGrandi2010_2}. Floquet APT (FAPT) 
extends this idea to a periodically driven system, which is relevant for our
setup with parameter $\lambda=A_{y}$ ramped slowly during
the pump pulse. Consider as before the case where the system starts
with drive turned off in the single particle eigenstate
$|0\rangle$ of undriven Hamiltonian $H(\lambda(t_{0}))$. Turning
on the drive slowly, the full time evolution is captured in the wave
function $|\psi(t)\rangle$. We can approximately solve the problem
by doing a unitary rotation to the moving frame: $|\tilde{\psi}\rangle=V^{\dagger}|\psi\rangle$,
where $V(\lambda(t),t)=P(\lambda,t)U_{d}(\lambda)$ is a unitary that
maps the Floquet eigenstates $|n_{F}(\lambda,t)\rangle$ (see Eq.
\ref{eq:floquet_eig_time_dep}) to a fixed basis $|e_{n}\rangle$.
In particular if we were to imagine turning on $\lambda$ infinitely
slowly in a gapped Floquet system, then the initial state $|0\rangle$
would just adiabatically track to the Floquet eigenstate $|\psi(t)\rangle=|0_{F}(\lambda,t)\rangle$
and $V^{\dagger}$ would act to map this to a time and $\lambda$-independent
state $|\tilde{\psi}\rangle=|e_{0}\rangle$. For a generic time evolution
$\lambda(t)$, the effective Hamiltonian in this moving frame is given
by
\begin{equation}
H_{m}=U_{d}^{\dagger}H_{F}U_{d}-i\dot{\lambda}V^{\dagger}\partial_{\lambda}V\equiv H_{F}^{d}-\dot{\lambda}\tilde{A}_{F},\label{eq:H_m_fapt}
\end{equation}
where $H_{F}^{d}$ is a diagonal matrix whose entries correspond to
the Floquet quasienenergies and $A_{F}(\lambda,t)=V\tilde{A}_{F}V^{\dagger}$
is the natural Floquet generalization of the Berry connection operator,
with matrix elements $\langle m_{F}(\lambda,t)|A_{F}|n_{F}(\lambda,t)\rangle=i\langle m_{F}(\lambda,t)|\partial_{\lambda}n_{F}(\lambda,t)\rangle$.
In the adiabatic limit ($\dot{\lambda}\to0$), off-diagonal elements
of the second term in Eq. \ref{eq:H_m_fapt} are unable to cause transitions,
which yields the adiabatic loading of the Floquet eigenstates as we
just discussed.

Floquet APT consists of solving leading corrections to adiabaticity
induced by the second term in Eq. \ref{eq:H_m_fapt}. As this term
is small due to the slow ramp rate $\dot{\lambda}$, it can be treated
perturbatively. In particular, one may note that at fixed $\lambda$,
$\tilde{A}_{F}$ is a periodic operator with Fourier series $\tilde{A}_{F}=\sum_{\ell}\tilde{A}_{F}^{(\ell)}e^{i\ell\Omega t}$
and similarly for $V$.
Then Eq. \ref{eq:H_m_fapt} yields a Floquet problem which we can
approximately solve using \emph{static} perturbation theory. Expanding
the wave function $|\psi(t)\rangle=\sum_{n}c_{n}|n_{F}(\lambda(t),t)\rangle$,
the coefficients at leading order in adiabatic perturbation theory
are given by \cite{Drese1999_1}
\begin{eqnarray}
\nonumber
c_{0} & \approx & e^{-i\Theta_{0}(t)}\\
c_{n} & \approx & e^{-i\Theta_{0}(t)}\dot{\lambda}(t)\sum_{\ell}\frac{\langle e_{n}|\tilde{A}_{F}^{(\ell)}(\lambda(t))|e_{0}\rangle}{\epsilon_{n}^{F}(\lambda)-\epsilon_{0}^{F}(\lambda)+\ell\Omega}e^{i\ell\Omega t}~.
\label{eq:coeff_fapt}
\end{eqnarray}
The phase $\Theta_{0}$ that the wave function picks up
during the ramp consists of a dynamical and a Berry phase: 
\begin{equation*}
\Theta_{0}(t)=\int_{t_{0}}^{t}\left[\epsilon_{0}^{F}(\lambda(t^\prime))-\dot{\lambda}(t^\prime)\langle e_{0}|\tilde{A}_{F}(\lambda(t^\prime),t^\prime)|e_{0}\rangle\right]dt^\prime.
\end{equation*}
This phase is usually neglected in most APT calculations of single-time
observables, but is crucial to situations like ARPES where non-equilibrium
observables are measured.

\subsection{Application of FAPT to Wigner distribution}

One can now use the approximate wave function $|\psi(t)\rangle$ derived above to obtain the Wigner distribution, $f(t_\mathrm{av},t_\mathrm{r})  =  \langle\psi(t_\mathrm{av}+t_\mathrm{r}/2)|\psi(t_\mathrm{av}-t_\mathrm{r}/2)\rangle$. For this Floquet problem, time enters in two ways:
in the periodic part of the Floquet eigenstates, and in the slow time-dependence of $\lambda$. In the spirit of FAPT, we expand this slow dependence about the measurement point $t_\mathrm{av}$, $\lambda(t_\mathrm{av} \pm t_\mathrm{r}/2)=\lambda(t_\mathrm{av})\pm t_\mathrm{r}\dot{\lambda}(t_\mathrm{av})/2+O(\dot{\lambda}^{2})$, and solve for the signal $I$ keeping all terms to order $\dot \lambda$. This calculation is done in detail in Appendix \ref{sec:fapt_details},
with the following result:
\begin{eqnarray}
I(\omega,t_\mathrm{pr})&\approx&\sum_{\ell}\left[(I^{(\ell)}_0+\Delta I^{(\ell)} ) e^{-[\omega-\omega_0^{(\ell)}-\Delta \omega^{(\ell)}]^{2}\tau_\mathrm{pr}^{2}}\right] \label{eq:I_om_tpr_f}\\
\nonumber
\Delta \omega^{(\ell)} & = & \dot \lambda \left( \partial_{\lambda}\varphi^{(\ell)}-\sum_{\ell^\prime}p_{0\ell^\prime}\partial_{\lambda}\varphi^{(\ell^\prime)} \right)\\
\nonumber
\frac{\Delta I^{(\ell)}}{I^{(\ell)}_0} & = & \dot \lambda \sum_{n,\ell^\prime}\Bigg(\frac{\langle e_{0}|V^{(-\ell)\dagger}V^{(-\ell-\ell^{\prime})}|e_{n}\rangle\langle e_{n}|\tilde{A}_{F}^{(\ell^{\prime})}|e_{0}\rangle}{\epsilon_{n0}^{F}+\ell^{\prime}\Omega}\Bigg)
\end{eqnarray}
where all expressions are evaluate at time $t_\mathrm{pr}$, the sum is taken for all pairs $(n,\ell^\prime) \neq (0,\ell)$, and notations are explained in the following paragraph.

\begin{figure}
\includegraphics[width=.6\columnwidth]{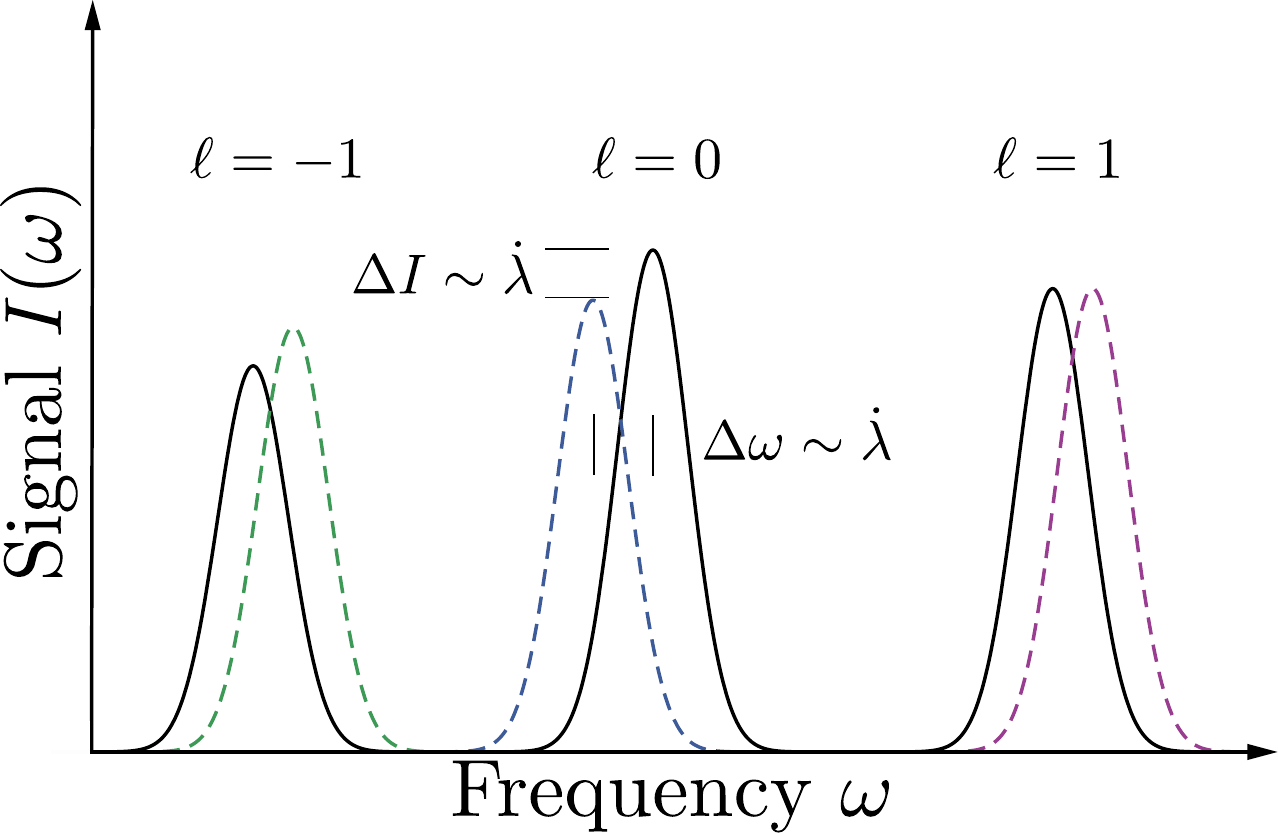}
\caption{\label{fig:fapt_arpes_signal}
Illustration of the effects of non-adiabaticity on the Floquet-ARPES signal with the FAPT approximation,
leading to shifts in both the peak frequency and height proportional to the ramp rate $\dot \lambda$.
}
\end{figure}

The effects of these leading corrections to adiabaticity
on the ARPES signal are illustrated in Fig. \ref{fig:fapt_arpes_signal}. 
Both the intensity $I^{(\ell)}_0=p_{0\ell}=\langle 0_F^{(\ell)}|0_F^{(\ell)}\rangle$ and the frequency $\omega_0^{(\ell)}=\epsilon_{0}^{F} - \ell\Omega$ of the Floquet sideband $|0_F^{(\ell)}\rangle$ defined in Eq.~\ref{eq:n_F_ell} are modified by an amount proportional to the ramp rate $\dot \lambda$. 
The intensity shift
$\Delta I^{(\ell)}$ results from virtual excitations
of $|0_{F}^{(\ell^{\prime})}\rangle$ to
$|n_{F}^{(\ell+\ell^{\prime})}\rangle$, which is a relatively standard prediction of adiabatic perturbation theory. Much more surprising are the frequency shifts, as they turn out to come from 
Berry phase effects. If we isolate the Berry phase sidebands as $|0_{F}^{(\ell)}\rangle=e^{i\varphi^{(\ell)}(\lambda)}|\tilde{0}_{F}^{(\ell)}(\lambda)\rangle$
such that $|\tilde{0}_{F}^{(\ell)}\rangle$ has vanishing Berry connection, then $\Delta \omega^{(\ell)}$
gives the difference of the Berry connection in sideband $\ell$ from
the average Berry connection across all sidebands. This object seems
somewhat bizarre if for no other reason than the fact that the Berry
connection is not gauge-invariant. However, this difference of Berry
connections \emph{is} gauge invariant and leads to a Berry phase-dependent
shift of the frequency of the sidebands.

Interestingly, while we think of adiabatic perturbation theory as primarily holding in the limit of small velocities, the results above actual hold in the limit of large (but not too large) velocities in which resonances can be neglected. Similar to the results found earlier in the resonant limit, these corrections in FAPT will lead to an asymmetry in the intensity signal with respect to time $t=0$, even though the Gaussian pulse is symmetric with respect to $t=0$. Unlike the resonant case, these corrections get smaller as the velocity decreases, or equivalently the pump time $\tau_\mathrm{pump}$ increases, and the excitations that they describe are virtual, meaning that no real population will remain in the bulk. Combining this with our previous results, we see that as $\tau_\mathrm{pump}$ is increased from zero, we get crossovers between various regimes, which are 
\begin{enumerate}
\item $\tau_\mathrm{pump} \ll 1/J,1/\Delta$: Non-universal physics related to microscopic details.
\item $1/J,1/\Delta \ll \tau_\mathrm{pump} \ll 1/\Delta_\mathrm{res}$: Virtual excitations described by Floquet adiabatic perturbation theory.
\item $1/\Delta_\mathrm{res} \ll \tau_\mathrm{pump}$: Real excitations due to surface-bulk resonances.
\end{enumerate}
In the low-frequency weak-drive limit, we expect these regimes to be well separated \cite{Rudner2013_1}, but whether such a separation of 
scales occurs in general is an important open question.

\section{Discussion and conclusions}
\label{conclusions}

We have computed the Wigner distribution function for a driven topological insulator with bulk- 
surface coupling and study the effects of a pump pulse that weakly breaks the periodicity.
If the drive is fixed, the Floquet states are well defined. However, the slow turn on and off of the 
drive breaks this periodicity and the Floquet states are no longer solutions of the Schr\"odinger equation.
This leads to non-adiabatic population transfer from the surface states
to the bulk. We track the origin to the existence of bulk-surface avoiding 
crossings in the the quasienergy spectrum, a signature of which are oscillations 
in the ARPES signal of a pump-probe type of experiment.

Finally we computed, using perturbation theory on the ramp rate of the drive amplitude, 
leading corrections to the ``adiabatic'' Floquet states. We showed that there are shifts of the 
resonances in the quasienergy spectrum. The shifts are a measure of the generalization of
the Berry connection to periodically driven systems and can theoretically be seen in the ARPES spectrum.
 
These novel surface-bulk coupling effects are a very interesting paradigm to explore in
future research. Many probes involve this basic setup, including ARPES, various types of scanning tip
microscopy, photon-in photon-out scattering experiments, and many others. In systems with 
interesting topological surface states, or even traditional non-topological ones, this bulk/surface
coupling upon resonant drive should yield interesting physically-measurable effects.

Topological insulators are rather weakly-correlated materials \cite{Xia2009_1,Hsieh2009_2}, so our treatment of them as 
non-interacting is well-justified. Generally, one expects this story to hold up against
weak experimental realities such as interactions or disorder as long as the timescales associated
with these processes are slower than those of the coherent bulk-surface oscillations. A more
experimentally-relevant concern are phonons, which generally have a much stronger effect
on bulk states than surface states \cite{Wang2012_1}. This could have the potentially
interesting effects of preferentially dephasing or relaxing higher harmonics of the surface states
due to the presence of nearby in energy bulk states coupled by bulk phonons, while having a much
weaker effect on surface harmonics that remain in the bulk gap. The effects of these experimentally-relevant
factor remains an open topic for future research.

Finally, we note that driving the surface states of TIs and other materials was spurred by
the search for novel topological states \cite{Oka2009_1,Kitagawa2011_1,Lindner2011_1} and rapidly expanded
to other contexts\cite{Grushin2014_1,Chan2016_1,Fruchart2016_1,Quelle2014_1,Choudhury2014_1,Dahlhaus2015_1,Perez-Piskunow2015_1,Sacramento2015_1,Liu2015_1,Privitera_1}. 
In particular, it was proposed that driving a Dirac cone by circularly-polarized light could open
a topological gap, yielding a Floquet Chern insulator. These proposals formally utilize the limit
where $\Omega$ is much larger than the band gap, but experiments practically work in the opposite limit.
The interesting open question is then what aspects of this topological character remain. There have
been a number of recent studies that explored the interplay of bulk and surface states in systems driven
at low frequencies \cite{Kitagawa2010_1,Rudner2013_1,TitumArxiv2015_1,Carpentier2015_1}, in which novel topological invariants were discovered 
that explicitly depend on the Floquet structure.
However, those papers consider bulk driving of an initially trivial system, whereas our paper
considers surface driving of an initially non-trivial system. We find seemingly unavoidable surface-bulk
coupling which seems to close the Floquet gap and break down this topological classification for such driving. 
However, topological protection can also extend to gapless systems \cite{Wan2011_1,Matsuura2013_1}, so 
we leave the open question of how this surface driving affects the topological classification of the TI
for future work.

\section{Acknowledgments}
We thank J. Freericks, N. Gedik, A. Kemper, F. Mahmood, T. Morimoto, and M. Sentef for useful discussions. 
BMF acknowledges support from AFOSR MURI, Conacyt, and computing resources from NERSC contract No. DE-AC02-05CH11231.
MK and JEM acknowledge support from
Laboratory Directed Research and Development (LDRD) funding
from Berkeley Lab, provided by the Director, Office
of Science, of the U.S. Department of Energy under Contract
No. DEAC02-05CH11231.


\begin{thebibliography}{103}
\bibitem [{Foo()}]{Footnote_ARPES_remove_res}%
  \BibitemOpen
  \href@noop {} {}\bibinfo {note} {To get the ``Floquet equilbrium'' data shown
  in the figures for the bulk/surface coupled model, we have manually removed
  the resonances at each value of $\vect k$. This is accomplished by solving
  the Floquet problem exactly for a few different values of $L$ between $85$
  and $115$. Then we select the value of $L$ for which the Floquet eigenstate
  has the most weight near the surface (i.e., for which $\overline{\langle
  z\rangle}$ is minimized). Each $L$ has a slightly different value of the
  resonances, so we are able to roughly get a smooth interpolations of the data
  ``without resonances.'' This interpolating curve between the many values of
  $L$ is picked out by eye and then fitted with a fifth-order
  polynomial.}\BibitemShut {Stop}%
\end{thebibliography}
%

\appendix
\onecolumngrid

\section{Further details of boundary-driven bulk TI model\label{sec:bulk_ti_model_details}}

In this appendix we briefly provide a more concrete definition of
the Hamiltonian described the main text. As mentioned earlier, $k_{x,y}$
are conserved quantities, while $k_{z}$-dispersion becomes hopping.
Labeling the sites along the $z$-direction as $j=0,1,\ldots,L-1$,
the Hamiltonian may then be written 
\begin{eqnarray*}
H & = & J \left( \begin{array}{cccc}
H_{\mathrm{d}}^{0} & H_{\mathrm{od}} & \\
H_{\mathrm{od}}^{\dagger} & H_{\mathrm{d}}^{1} & H_{\mathrm{od}} &  \\
 & H_{\mathrm{od}}^{\dagger} & \ddots & \\
 &  &  & H_{d}^{L-1}
\end{array} \right) ,\\
H_{\mathrm{d}}^{j} & = & \tau^{z}\left[\sigma^{x}\sin(k_{x}+a_{x}^{j})+\sigma^{y}\sin(k_{y}+a_{y}^{j})\right]+\tau^{x}\left[m+3-\cos(k_{x}+a_{x}^{j})-\cos(k_{y}+a_{y}^{j})\right],\\
H_{\mathrm{od}} & = & \frac{i\sigma^{z}\tau^{z}-\tau^{x}}{2},
\end{eqnarray*}
where the position-dependent vector potentials are $a_{x}^{j}=A_{x}\sin(\Omega t)\exp(-j/\xi)$
and $a_{y}^{j}=A_{y}\cos(\Omega t)\exp(-j/\xi)$. As noted in the
main text, we work in the case $A_{x}=0$ and $k_{y}=0$ for all of the data
shown.

\section{Further details of the Demkov-Osherov model\label{sec:demkov_osherov_details}}

The Demkov-Osherov (D-O) model consists of $N$ parallel levels traversed
by a single mode whose energy changes linearly with some parameter
$\lambda$ \cite{Demkov1968_1,Macek1998_1,Sinitsyn2002_1}. It can formally be solved when treated
as a scattering problem, i.e., starting with some probability $p_{n}^{i}$
in the states at $\lambda(t=-\infty)=-\infty$, $\lambda$ is ramped
linearly according to $\lambda=vt$ and the final probabilities at
$t=\infty$ are obtained. The nice property of this model is that
the level-crossings factorize, in the sense that the probability of
ending up in one branch can be obtained by simply taking the semi-classical
product of all the prior two-level (Landau-Zener) avoided crossings.
Essentially this implies that in the long-time limit there are no
interference effects between the various avoided crossings.

\begin{figure}
\includegraphics[width=0.7\textwidth]{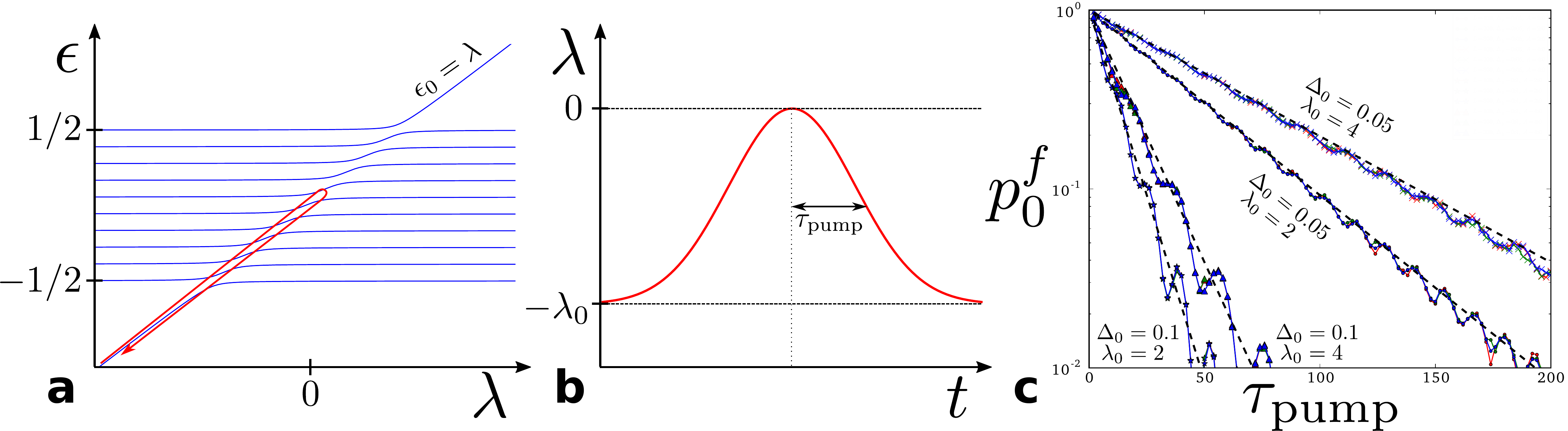}

\caption{\label{fig:demkov_osherov_supplement}Surface state occupation in
Gaussian ramps of the Demkov-Osherov (D-O) model. (a) Illustration
of the simplified D-O model that we consider (Eq.~\ref{eq:H_do})
with $L=10$ levels. We simulate a Gaussian ramp to the middle of
the bulk spectrum (b) and track the final surface state occupation
$p_{0}^{f}=|\langle0|\psi(t=\infty)\rangle|^{2}$ as function of the
ramp rate $\tau_{\mathrm{pump}}$ (c). The dashed lines show the predicted
value $p_{0}^{f}=e^{-2\Gamma\tau_{\mathrm{\mathrm{pump}}}}$ with
$\Gamma$ from Eq.~\ref{eq:Gamma_gaussian} showing a good fit for
a variety of $\Delta_{0}$ and $\lambda_{0}$.}
\end{figure}

Motivated by the surface-bulk resonance discussed in Sec.~\ref{sec:floquet_res_lz}, 
we will consider a particular subclass of D-O model illustrated
in Fig.~\ref{fig:demkov_osherov_supplement}a. A total of $L$ levels
representing the bulk bands span the energy window $\epsilon_{\mathrm{bulk}}\in(-1/2,1/2)$
while the surface state disperses with bare energy $\epsilon_{0}=\lambda$
with some generic parameter $\lambda$ taking the place of $A_{y}$.
Gaps of strength $2\Delta_{0}/\sqrt{L}$ are opened uniformly between
each bulk state and the surface state, which we will see gives a well-defined
thermodynamic limit. Choosing all matrix elements to be real and labeling
the bulk states $|j=1,\cdots,L\rangle$ and the surface state $|0\rangle$,
this is described by the Hamiltonian 
\begin{equation}
H=\lambda|0\rangle\langle0|+\sum_{j=1}^{L}\epsilon_{j}|j\rangle\langle j|+\frac{\Delta_{0}}{\sqrt{L}}\sum_{j=1}^{L}\left(|0\rangle\langle j|+|j\rangle\langle0|\right),\label{eq:H_do}
\end{equation}
where $\epsilon_{j}=(j-1/2)/L-1/2$.

We will be particularly interested in taking this model to the thermodynamic
limit $L\to\infty$ and ascertaining what universal properties can
be found in its dynamics. Consider first the exactly-solvable case
where we start in the ground state $|0\rangle$ at $\lambda=-\infty$
and ramp linearly via $\lambda=vt$. Due to the fact that the crossings
can be treated independently, at time $t=\infty$ the probability
to remain in the surface state is simply $p_{0}^{f}=\prod_{j}e^{-2\pi\Delta_{j}^{2}/v}=e^{-2\pi\Delta_{0}^{2}/v}$,
where $\Delta_{j}=\Delta_{0}/\sqrt{L}$ is the off-diagonal matrix
element between $|0\rangle$ and $|j\rangle$. Note that this transition
probability is identical to that of a single Landau-Zener transition
with matrix element $\Delta_{\mathrm{eff}}=\Delta_{0}$. While this
effective gap only formally gives the final transition amplitude,
one can readily confirm numerically that the dynamics of the occupation
$p_{0}(t)=|\langle0|\psi(t)\rangle|^{2}$ during the ramp is also
well-approximated by that of a single avoided crossing of strength
$\Delta_{\mathrm{eff}}$. 

Let us now apply this intuitive approximation to arbitrary ramps $\lambda(t/\tau_{\mathrm{pump}})$
set by some timescale $\tau_{\mathrm{pump}}$ (e.g., the width of
a Gaussian). A natural estimate for the transition probabilities is
that they will again factorize but now with $v\to|v_{j}|=|\dot{\epsilon}_{0}(t_{j})|$,
where $t_{j}$ is the time where the $j$th level is crossed: $\epsilon_{0}(\lambda(t_{j}))=\epsilon_{j}$.
Then if we start in the state $|0\rangle$ at time $t_{i}$ and monotonically
increase $\lambda$ up to time $t_{f}$, such that $|v_{j}|=v_{j}$,
the amount remaining in the surface state will be $p_{0}^{t_{i}\to t_{f}}\approx e^{-\alpha(t_{i}\to t_{f})}$,
where
\begin{equation}
\alpha(t_{i}\to t_{f})=2\pi\sum_{i}\Delta_{i}^{2}/v_{i}\stackrel{L\to\infty}{\rightarrow}2\pi\int d\epsilon\nu(\epsilon)\frac{\Delta^{2}(\epsilon)}{d\epsilon/dt}=2\pi\Delta_{0}^{2}\int_{t_{1}}^{t_{f}}dt=2\pi\Delta_{0}^{2}|z_{1}|\tau_{\mathrm{pump}},\label{eq:alpha_eff_ramp_up}
\end{equation}
$\nu(\epsilon)=L$ is the density of bulk states, and $t_{1}(\lambda_{i},\lambda_{f})=z_{1}(\lambda_{i},\lambda_{f})\tau_{\mathrm{pump}}$
is the time where the surface state first passes into the bulk, i.e.,
where it crosses $\epsilon_{1}$. Note that this can be written as
$p_{0}^{t_{i}\to t_{f}}\approx e^{-\Gamma\tau_{\mathrm{pump}}}$ which
looks like a constant rate $\Gamma$ of surface states leaking into
the bulk during the ramp.

The story becomes even more subtle if $\lambda(t)$ is not monotonic.
Then the surface state may cross a given bulk state multiple times,
and population that had transferred into the bulk may now return to
the surface. However, we are already ignoring interference effects
in the above model by, for instance, not ramping all the way to $\lambda=\infty$
to dephase the excitations. Therefore, at a similar level of approximation
we may assume that no population, once transferred to the bulk, is
able to return to the surface. Furthermore, if $\lambda(t)$ is an
even function of time, then the magnitude of the velocity $v_{j}$
for passing bulk level $j$ during the first half of the ramp will
be the same as during the second half of the ramp. Thus, we estimate
the final surface occupation to be $p_{0}^{f}=e^{-2\Gamma\tau_{\mathrm{pump}}}$,
where $\Gamma$ is given by Eq.~\ref{eq:alpha_eff_ramp_up}. We numerically
test this approximation using a Gaussian ramp that starts from $\lambda=-\lambda_{0}$
and ramps to $\lambda=0$ as illustrated in Fig.~\ref{fig:demkov_osherov_supplement}b.
Plugging this ramp profile into Eq.~\ref{eq:alpha_eff_ramp_up},
we find
\begin{equation}
\Gamma_{\mathrm{Gaussian}}=2\pi\Delta_{0}^{2}\sqrt{-2\ln(1-1/(2\lambda_{0}))}.\label{eq:Gamma_gaussian}
\end{equation}
This estimate is plotted against exact simulation in Fig.~\ref{fig:demkov_osherov_supplement}c,
showing a good fit. This justifies our independent-level Demkov-Osherov
approximation for Gaussian ramps, which is used in the main text to
fit the data in Fig.~\ref{fig:signal_surface_bulk_kcut_change_params}.

\section{Matrix elements in ARPES\label{sec:matrix_elements_arpes}}

An additional complication in interpreting ARPES experiments is the fact that
not all electrons photoemit with identical matrix elements, as we have tacitly assumed throughout this work.
The general expression for the ARPES signal in the presence
of photoemission matrix elements is significantly more complicated
\cite{Freericks2009_1} and does not provide much insight to our analysis. However, we can slightly improve our approximation
by simply weighting the states in $G^{<}$ by their position along
the $z$-direction. The intuition behind this is that both the probe
photons and the ionized electrons have some finite penetration depth
or mean free path in the bulk before they are dissipated. Approximating
this by a single length scale $\xi_\mathrm{pr}$, we can introduce a weighting
operator
\[
\hat{W}_{\vect k_\parallel}=\sum_{j\alpha}e^{-j/\xi_\mathrm{pr}}|j\alpha\vect k_\parallel\rangle\langle j\alpha\vect k_\parallel|,
\label{eq:arpes_matrix_elem}
\]
where $j=0,1,\ldots,L-1$ is the site number along the $z$-direction,
$\alpha=1-4$ are indices in the spin-orbital basis of $\sigma$ and
$\tau$, and $\vect k_\parallel$ is the $xy$ momentum as before. This operator
just weights single-particle states by their position along $z$ and
thus we approximate the surface-weighted ARPES response by replacing
$f$ by 
\[
f_{n}^{\prime}(t_\mathrm{r},t_\mathrm{av})=\langle\psi_{n}(t_\mathrm{av}+t_\mathrm{r}/2)|\hat{W}|\psi_{n}(t_\mathrm{av}-t_\mathrm{r}/2)\rangle.
\]

\begin{figure}
\includegraphics[width=.5\columnwidth]{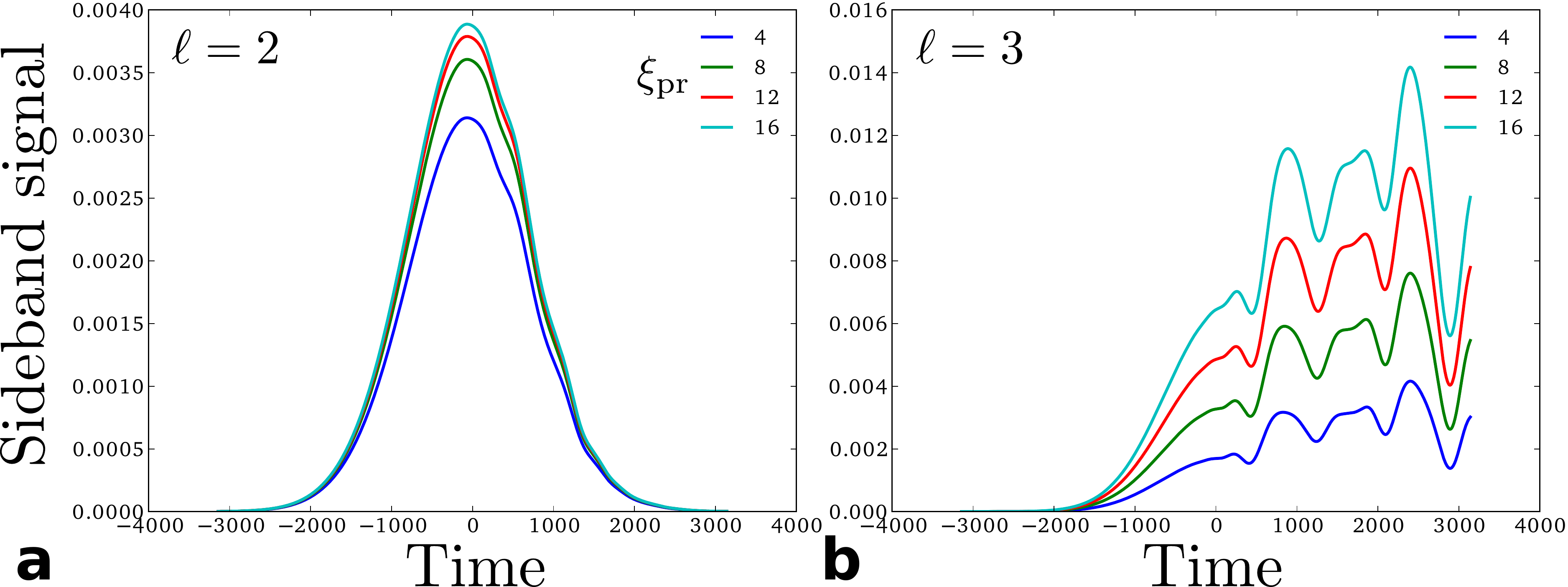}
\caption{\label{fig:arpes_with_matrix_elem}
Signal in the $\ell=2$ and $3$ sidebands as a function of the length scale $\xi_\mathrm{pr}$ for
approximate ARPES matrix elements (Eq.~\ref{eq:arpes_matrix_elem}) using the same
parameters as Fig.~\ref{fig:driven_ss_plus_bulk}b. Data in previous figures is essentially
the $\xi_\mathrm{pr}\to\infty$ limit of this model.
}
\end{figure}

The result with this surface projection are shown in Fig. \ref{fig:arpes_with_matrix_elem}
and allow us to compare surface and bulk behavior, particularly in higher Floquet sidebands.
We see that the $\ell=2$ sideband does not change significantly in either amplitude or character
as $\xi_\mathrm{pr}$ is varied, which is consistent with its nature as a surface state. On the other hand,
the $\ell=3$ sideband is dominated by excitations into the bulk, which shows up as a strong
increase in the signal with $\xi_\mathrm{pr}$. On top of these bulk excitations, one expects a surface
sideband signal as well, which should not depend on $\xi_\mathrm{pr}$ in the $\xi_\mathrm{pr} \to \infty$
limit. In principle we should be able to use this idea to distinguish the surface and bulk signals.
Unfortunately, we are currently unable to do so with our data due to finite size effects; we leave
this distinction of surface and bulk signals in the sidebands as a subject for future work.

\section{Further details of FAPT\label{sec:fapt_details}}

In this appendix, we will derive Eq.~\ref{eq:I_om_tpr_f} by using the approximate time-dependent wave function derived using FAPT (Eq.~\ref{eq:coeff_fapt}) to obtain the Wigner distribution:
\begin{eqnarray}
f(t_\mathrm{av},t_\mathrm{r}) & = & \langle\psi(t_\mathrm{av}+t_\mathrm{r}/2)|\psi(t_\mathrm{av}-t_\mathrm{r}/2)\rangle\equiv\langle\psi(t_{+})|\psi(t_{-})\rangle\nonumber \\
 & \approx & e^{-i(\Theta_{0}(t_{-})-\Theta_{0}(t_{+}))}\Bigg[\langle0_{F}(\lambda_{+},t_{+})|0_{F}(\lambda_{-},t_{-})\rangle+ \nonumber \\ &&\dot{\lambda}_{-}\sum_{n\neq0,\ell}\frac{\langle e_{n}|\tilde{A}_{F}^{(\ell)}(\lambda_{-})|e_{0}\rangle}{\epsilon_{n0}^{F}(\lambda_{-})+\ell\Omega}e^{i\ell(\Omega t_{-}-\varphi_{0})}
 \langle0_{F}(\lambda_{+},t_{+})|n_{F}(\lambda_{-},t_{-})\rangle+(\lambda_{+}\leftrightarrow\lambda_{-})\Bigg],\label{eq:f_orig}
\end{eqnarray}
where $t_{\pm}\equiv t_\mathrm{av}\pm t_\mathrm{r}/2$ and $\lambda_\alpha \equiv \lambda(t_\alpha)$. As mentioned in the main text, time enters via both the periodic part of the Floquet eigenstates and the slow time-dependence
of $\lambda$, and we will expand this slow dependence about the $t_\mathrm{av}$: $\lambda(t_{\pm})=\lambda(t_\mathrm{av})\pm t_\mathrm{r}\dot{\lambda}(t_\mathrm{av})/2+O(\dot{\lambda}^{2})$.

Let us now evaluate the terms in Eq. \ref{eq:f_orig} one-by-one. First,
consider the phase factor $e^{i\delta\Theta_{0}}$ where 
\begin{equation}
\delta\Theta_{0}=\Theta_{0}(t_{+})-\Theta_{0}(t_{-})=\int_{t_{-}}^{t_{+}}\left(\epsilon_{0}^{F}(\lambda(t^\prime))-\dot{\lambda}(t^\prime)\langle e_{0}|\tilde{A}_{F}(\lambda(t^\prime),t^\prime)|e_{0}\rangle\right)dt^\prime~.\label{eq:delta_Theta_0}
\end{equation}
At order $\dot{\lambda}$, the energy can be expanded around $\lambda_\mathrm{av}$
as $\epsilon_{0}^{F}(\lambda(t^\prime))\approx\epsilon_{0}^{F}(\lambda_\mathrm{av})+\dot{\lambda}_\mathrm{av}(t^\prime-t_\mathrm{av})\partial_{\lambda}\epsilon_{0}^{F}(\lambda_\mathrm{av})$.
The second term is odd about $t_\mathrm{av}$, so it integrates to zero.
Meanwhile, it is useful to express $\tilde{A}_{F}$ and $V$ in terms
of Fourier modes: 
\begin{eqnarray}
V(\lambda,t) & = & \sum_{\ell}e^{i\ell(\Omega t-\varphi_{0})}V^{(\ell)}(\lambda)\\
\tilde{A}_{F}(\lambda,t) & = & i\sum_{\ell,\ell^\prime}e^{-i\ell^\prime(\Omega t-\varphi_{0})}V^{(\ell^\prime)\dagger}\partial_{\lambda}V^{(\ell)}e^{i\ell(\Omega t-\varphi_{0})}\\
\implies\tilde{A}_{F}^{(\ell)} & = & i\sum_{\ell^\prime}V^{(\ell^\prime)\dagger}\partial_{\lambda}V^{(\ell+\ell^\prime)}.
\end{eqnarray}
Throughout this appendix, we explicitly write the driving phase $\varphi_0$ to facilitate averaging over it as in Eq.~\ref{eq:f_n_omega_bar}.
Then, replacing $\lambda(t^\prime)$ by $\lambda_\mathrm{av}$ in the second term
of Eq. \ref{eq:delta_Theta_0} to leading order in $\dot{\lambda}$
we get 
\begin{eqnarray}
\int_{t_{-}}^{t_{+}}\langle e_{0}|\tilde{A}_{F}(\lambda_\mathrm{av},t^\prime)|e_{0}\rangle dt^\prime & = & \sum_{\ell}\int_{t_{-}}^{t_{+}}e^{i\ell(\Omega t^\prime-\varphi_{0})}\langle e_{0}|A_{F}^{(\ell)}|e_{0}\rangle dt^\prime\\
 & = & \sum_{\ell}\frac{\langle e_{0}|\tilde{A}_{F}^{(\ell)}|e_{0}\rangle}{i\ell\Omega}\left(e^{i\ell(\Omega t_{+}-\varphi_{0})}-e^{i\ell(\Omega t_{-}-\varphi_{0})}\right)\\
 & = & \sum_{\ell}\frac{\langle e_{0}|\tilde{A}_{F}^{(\ell)}|e_{0}\rangle}{i\ell\Omega}e^{i\ell(\Omega t_\mathrm{av}-\varphi_{0})}\left(e^{i\ell\Omega t_\mathrm{r}/2}-e^{-i\ell\Omega t_\mathrm{r}/2}\right)\equiv iB_{1}~.
\end{eqnarray}
Unless explicitly stated otherwise, all terms in the above expression
are now evaluated at $\lambda_\mathrm{av}$, which is a trick we will employ
throughout. Putting these terms together, 
\begin{equation}
e^{i\delta\Theta_{0}}\approx e^{i\epsilon_{0}^{F}t_\mathrm{r}}e^{-i\dot{\lambda}_\mathrm{av}(iB_{1})}\approx e^{i\epsilon_{0}^{F}t_\mathrm{r}}\left(1+\dot{\lambda}_\mathrm{av}B_{1}\right)~.
\end{equation}
Note that this first term in this product gives the main peak center
as $\epsilon_{0}^{F}$, the Floquet quasi-energy, while the terms
like $e^{i\ell\Omega t_\mathrm{r}/2}$ in $B_{1}$ give additional satellite
peaks offset by half-integer multiples of $\Omega$. Later averaging
over the phase $\varphi_{0}$ will remove all but the integer multiples of this frequency.

Next, we Taylor expand the term that appears to be $O(\dot{\lambda}^{0})$
in Eq. \ref{eq:f_orig} about time $t_\mathrm{av}$: 
\begin{equation}
|0_{F}(\lambda(t_{-}),t_{-})\rangle\approx|0_{F}(\lambda_\mathrm{av},t_{-})\rangle-\frac{t_\mathrm{r}}{2}\dot{\lambda}_\mathrm{av}\partial_{\lambda}|0_{F}(\lambda_\mathrm{av},t_{-})\rangle
\end{equation}
and similarly for the bra. Thus, 
\begin{equation}
\langle0_{F}(\lambda_{+},t_{+})|0_{F}(\lambda_{-},t_{-})\rangle\approx\underbrace{\langle0_{F}(\lambda_\mathrm{av},t_{+})|0_{F}(\lambda_\mathrm{av},t_{-})\rangle}_{A_{0}}+\underbrace{\frac{\dot{\lambda}_\mathrm{av}t_\mathrm{r}}{2}\left(\langle\partial_{\lambda}0_{F}(\lambda_\mathrm{av},t_{+})|0_{F}(\lambda_\mathrm{av},t_{-})\rangle-\langle0_{F}(\lambda_\mathrm{av},t_{+})|\partial_{\lambda}0_{F}(\lambda_\mathrm{av},t_{-})\rangle\right)}_{\dot{\lambda}A_{1}}.
\end{equation}
Now $|0_{F}(\lambda,t)\rangle=V(\lambda,t)|e_{0}\rangle$, so $\partial_{\lambda}|0_{F}\rangle=\partial_{\lambda}V|e_{0}\rangle$.
Thus 
\begin{eqnarray}
A_{1} & = & \frac{t_\mathrm{r}}{2}\left(\langle e_{0}|\partial_{\lambda}V^{\dagger}(t_{+})V(t_{-})|e_{0}\rangle-\langle e_{0}|V^{\dagger}(t_{+})\partial_{\lambda}V(t_{-})|e_{0}\rangle\right)\\
 & = & \frac{t_\mathrm{r}}{2}\sum_{\ell^\prime,\ell^{\prime \prime}}\left(\langle e_{0}|\partial_{\lambda}V^{(\ell^\prime)\dagger}e^{-i\ell^\prime(\Omega t_{+}-\varphi_{0})}e^{i\ell^{\prime\prime}(\Omega t_{-}-\varphi_{0})}V^{(\ell^{\prime\prime})}|e_{0}\rangle-\langle e_{0}|V^{(\ell^{\prime})\dagger}e^{-i\ell^{\prime}(\Omega t_{+}-\varphi_{0})}e^{i\ell^{\prime\prime}(\Omega t_{-}-\varphi_{0})}\partial_{\lambda}V^{(\ell^{\prime\prime})}|e_{0}\rangle\right)\\
 & = & \frac{t_\mathrm{r}}{2}\sum_{\ell^{\prime},\ell^{\prime\prime}}e^{-i(\ell^{\prime}-\ell^{\prime\prime})(\Omega t_\mathrm{av}-\varphi_{0})}e^{-i(\ell^{\prime\prime}+\ell^{\prime})\Omega t_\mathrm{r}/2}\left(\langle e_{0}|\partial_{\lambda}V^{(\ell^{\prime})\dagger}V^{(\ell^{\prime\prime})}|e_{0}\rangle-\langle e_{0}|V^{(\ell^{\prime})\dagger}\partial_{\lambda}V^{(\ell^{\prime\prime})}|e_{0}\rangle\right).
\end{eqnarray}
Meanwhile, 
\begin{equation}
A_{0}=\sum_{\ell^{\prime},\ell^{\prime\prime}}e^{-i(\ell^{\prime}-\ell^{\prime\prime})(\Omega t_\mathrm{av}-\varphi_{0})}e^{-i(\ell^{\prime\prime}+\ell^{\prime})\Omega t_\mathrm{r}/2}\langle e_{0}|V^{(\ell^{\prime})\dagger}V^{(\ell^{\prime\prime})}|e_{0}\rangle~.
\end{equation}

In the remaining two terms of Eq. \ref{eq:f_orig}, at order $\dot{\lambda}$
we can again replace $\lambda_{\pm}$ by $\lambda_\mathrm{av}$. Then we
can group these two terms into one that we denote $\dot{\lambda}_\mathrm{av}A_{2}$,
with 
\begin{eqnarray}
A_{2} & = & \sum_{n\neq0,\ell}\Big[\frac{\langle e_{n}|\tilde{A}_{F}^{(\ell)}|e_{0}\rangle}{\epsilon_{n0}^{F}+\ell\Omega}e^{i\ell(\Omega t_\mathrm{av}-\varphi_{0})}e^{-i\ell\Omega t_\mathrm{r}/2}\langle0_{F}(\lambda_\mathrm{av},t_{+})|n_{F}(\lambda_\mathrm{av},t_{-})\rangle+\\
 &  & ~~~~~~~\frac{\langle e_{0}|\tilde{A}_{F}^{(\ell)\dagger}|e_{n}\rangle}{\epsilon_{n0}^{F}+\ell\Omega}e^{-i\ell(\Omega t_\mathrm{av})-\varphi_{0}}e^{-i\ell\Omega t_\mathrm{r}/2}\langle n_{F}(\lambda_\mathrm{av},t_{+})|0_{F}(\lambda_\mathrm{av},t_{-})\rangle\Big].
\end{eqnarray}
Now 
\begin{eqnarray}
\langle0_{F}(t_{+})|n_{F}(t_{-})\rangle & = & \langle e_{0}|V(t_{+})^{\dagger}V(t_{-})|e_{n}\rangle\\
 & = & \langle e_{0}|\left(\sum_{\ell^{\prime},\ell^{\prime\prime}}V^{(\ell^{\prime})\dagger}e^{-i\ell^{\prime}(\Omega t_{+}-\varphi_{0})}e^{i\ell^{\prime\prime}(\Omega t_{-}-\varphi_{0})}V^{(\ell^{\prime\prime})}\right)|e_{n}\rangle\\
 & = & \langle e_{0}|\left(\sum_{\ell^{\prime},\ell^{\prime\prime}}e^{i(\ell^{\prime\prime}-\ell^{\prime})(\Omega t_\mathrm{av}-\varphi_{0})}e^{-i(\ell^{\prime\prime}+\ell^{\prime})\Omega t_\mathrm{r}/2}V^{(\ell^{\prime})\dagger}V^{(\ell^{\prime\prime})}\right)|e_{n}\rangle
\end{eqnarray}
and similarly for $\langle n_{F}(t_{+})|0_{F}(t_{-})\rangle$. Thus
\begin{eqnarray}
A_{2} & = & \sum_{n\neq0,\ell,\ell^{\prime},\ell^{\prime\prime}}\Big[\frac{\langle e_{n}|\tilde{A}_{F}^{(\ell)}|e_{0}\rangle\langle e_{0}|V^{(\ell^{\prime})\dagger}V^{(\ell^{\prime\prime})}|e_{n}\rangle}{\epsilon_{n0}^{F}+\ell\Omega}e^{i(\ell+\ell^{\prime\prime}-\ell^{\prime})(\Omega t_\mathrm{av}-\varphi_{0})}e^{-i(\ell+\ell^{\prime}+\ell^{\prime\prime})\Omega t_\mathrm{r}/2}+\\
 &  & ~~~~~~~~~~~~~\frac{\langle e_{0}|\tilde{A}_{F}^{(\ell)\dagger}|e_{n}\rangle\langle e_{n}|V^{(\ell^{\prime\prime})\dagger}V^{(\ell^{\prime})}|e_{0}\rangle}{\epsilon_{n0}^{F}+\ell\Omega}e^{-i(\ell+\ell^{\prime\prime}-\ell^{\prime})(\Omega t_\mathrm{av}-\varphi_{0})}e^{-i(\ell+\ell^{\prime}+\ell^{\prime\prime})\Omega t_\mathrm{r}/2}\Big].
\end{eqnarray}

Altogether, 
\begin{equation}
f\approx e^{i\epsilon_{0}^{F}t_\mathrm{r}}(1+\dot{\lambda}_\mathrm{av}B_{1})(A_{0}+\dot{\lambda}_\mathrm{av}A_{1}+\dot{\lambda}_\mathrm{av}A_{2})\approx e^{i\epsilon_{0}^{F}t_\mathrm{r}}(A_{0}+\dot{\lambda}_\mathrm{av}(\underbrace{A_{1}+A_{2}+A_{0}B_{1}}_{A_{3}})),
\end{equation}
where we can rewrite $A_{0}B_{1}$ as 
\begin{equation}
A_{0}B_{1}=\sum_{\ell,\ell^{\prime},\ell^{\prime\prime}}\frac{\langle e_{0}|\tilde{A}_{F}^{(\ell)}|e_{0}\rangle\langle e_{0}|V^{(\ell^{\prime})\dagger}V^{\ell^{\prime\prime}}|e_{0}\rangle}{\ell\Omega}e^{-i(\ell^{\prime}-\ell^{\prime\prime}-\ell)(\Omega t_\mathrm{av}-\varphi_{0})}e^{-i(\ell^{\prime\prime}+\ell^{\prime})\Omega t_\mathrm{r}/2}\left(e^{-i\ell\Omega t_\mathrm{r}/2}-e^{i\ell\Omega t_\mathrm{r}/2}\right)~.
\end{equation}
Together with the expressions for $A_{0,1,2}$ above, this is the
leading correction to $f(t_\mathrm{av},t_\mathrm{r})$. However, the observable
ARPES signal comes from Fourier transforming this to get $f(t_\mathrm{av},\omega)$,
then convolving in both the frequency and time direction by the Gaussian
probe of width $\tau_\mathrm{pr}$, $e^{-\omega^{2}\tau_\mathrm{pr}^{2}}$ and $e^{-(t_\mathrm{av}-t_\mathrm{pr})^{2}/\tau_\mathrm{pr}^{2}}$
respectively, to get the ARPES signal $I(t_\mathrm{pr},\omega)$ at frequency
$\omega$ for a probe centered at time $t_\mathrm{pr}$. In the limit $\tau_\mathrm{pr}\gg T$
this convolution averages over many cycles as discussed earlier, which we treat by
averaging over $\varphi_{0}$. Then, for instance, the ``adiabatic'' signal reduces to 
\begin{equation}
\overline{A_{0}}=\sum_{\ell}\langle e_{0}|V^{(\ell)\dagger}V^{(\ell)}|e_{0}\rangle e^{-i\ell\Omega t_\mathrm{r}}~,\label{eq:A_0_fns}
\end{equation}
which yields the same Wigner distribution as Eq.~\ref{eq:f_n_omega_bar}. 

Let's now calculate the leading correction, $\overline{A_{3}}=\overline{A_{1}}+\overline{A_{2}}+\overline{A_{0}B_{1}}$,
term by term: 
\begin{eqnarray}
\overline{A_{1}} & = & \frac{t_\mathrm{r}}{2}\sum_{\ell}e^{-i\ell\Omega t_\mathrm{r}}\left[\langle e_{0}|\partial_{\lambda}V^{(\ell)\dagger}V^{(\ell)}|e_{0}\rangle-\langle e_{0}|V^{(\ell)\dagger}\partial_{\lambda}V^{(\ell)}|e_{0}\rangle\right]\\
\overline{A_{2}} & = & \sum_{n\neq0,\ell^{\prime},\ell^{\prime\prime}}e^{-i\ell^{\prime}\Omega t_\mathrm{r}}\Big[\frac{\langle e_{n}|\tilde{A}_{F}^{(\ell^{\prime}-\ell^{\prime\prime})}|e_{0}\rangle\langle e_{0}|V^{(\ell^{\prime})\dagger}V^{(\ell^{\prime\prime})}|e_{n}\rangle}{\epsilon_{n0}^{F}+(\ell^{\prime}-\ell^{\prime\prime})\Omega}+h.c.\Big]\\
\overline{A_{0}B_{1}} & = & \sum_{\ell^{\prime},\ell^{\prime\prime}}\frac{\langle e_{0}|\tilde{A}_{F}^{(\ell^{\prime}-\ell^{\prime\prime})}|e_{0}\rangle\langle e_{0}|V^{(\ell^{\prime})\dagger}V^{(\ell^{\prime\prime})}|e_{0}\rangle}{(\ell^{\prime}-\ell^{\prime\prime})\Omega}\left(e^{-i\ell^{\prime}\Omega t_\mathrm{r}}-e^{-i\ell^{\prime\prime}\Omega t_\mathrm{r}}\right)~.
\end{eqnarray}
It is worth noting that $\overline{A_{0}B_{1}}$ naturally breaks
up into ``diagonal'' and ``off-diagonal'' terms corresponding
to $\ell^{\prime}=\ell^{\prime\prime}$ and $\ell^{\prime}\neq\ell^{\prime\prime}$ respectively. The diagonal
term can be rewritten as 
\begin{eqnarray}
\left(\overline{A_{0}B_{1}}\right)_{d} & = & \sum_{\ell^{\prime}=\ell^{\prime\prime}}\frac{\langle e_{0}|\tilde{A}_{F}^{(0)}|e_{0}\rangle\langle e_{0}|V^{(\ell^{\prime})\dagger}V^{(\ell^{\prime})}|e_{0}\rangle}{(\ell^{\prime}-\ell^{\prime\prime})\Omega}e^{-i\ell^{\prime}\Omega t_\mathrm{r}}\left(1-e^{-i(\ell^{\prime\prime}-\ell^{\prime})\Omega t_\mathrm{r}}\right)\\
 & = & -it_\mathrm{r}\langle e_{0}|\tilde{A}_{F}^{(0)}|e_{0}\rangle\sum_{\ell^{\prime}}\langle e_{0}|V^{(\ell^{\prime})\dagger}V^{(\ell^{\prime})}|e_{0}\rangle.
\end{eqnarray}
This term along with $\overline{A_{1}}$ are the only ones proportional
to $t_\mathrm{r}$. They actually give rise to a shift of the peaks, since
the Fourier transform of $it_\mathrm{r}e^{iEt_\mathrm{r}}$ is the derivative of
the delta function, $\delta^{\prime}(\omega-E)$. Combining these two terms
gives 
\begin{equation}
\overline{A_{1}}+\left(\overline{A_{0}B_{1}}\right)_{d}=t_\mathrm{r}\sum_{\ell}e^{-i\ell\Omega t_\mathrm{r}}\left[\frac{\langle e_{0}|\partial_{\lambda}V^{(\ell)\dagger}V^{(\ell)}|e_{0}\rangle-h.c.}{2}-i\langle e_{0}|\tilde{A}_{F}^{(0)}|e_{0}\rangle\langle e_{0}|V^{(\ell)\dagger}V^{(\ell)}|e_{0}\rangle\right].\label{eq:diff_gs_conn}
\end{equation}
At this point it is useful to introduce the notation $|n_{F}^{(\ell)}\rangle=V^{(\ell)}|e_{n}\rangle$
as the $\ell$-th Fourier mode of the $n$-th Floquet eigenstate as in Eq.~\ref{eq:n_F_ell}.
Then the first term in Eq. \ref{eq:diff_gs_conn} looks like the Berry
connection of $|0_{F}^{(\ell)}\rangle$ with the caveat that the state
is not normalized. More explicitly, if we make so local gauge choice
of states $|\tilde{0}_{F}^{(\ell)}(\lambda)\rangle$ such that their
Berry connection is zero, i.e., $\langle\tilde{0}_{F}^{(\ell)}(\lambda)|\partial_{\lambda}\tilde{0}_{F}^{(\ell)}(\lambda)\rangle=0$,
then rewriting $|0_{F}^{(\ell)}\rangle=e^{i\varphi^{(\ell)}(\lambda)}|\tilde{0}_{F}^{(\ell)}\rangle$
we find $\langle0_{F}^{(\ell)}|\partial_{\lambda}0_{F}^{(\ell)}\rangle=i\partial_{\lambda}\varphi^{(\ell)}\langle0_{F}^{(\ell)}|0_{F}^{(\ell)}\rangle= i\partial_{\lambda}\varphi^{(\ell)}p_{0\ell}$.
Factoring this out of each term in Eq. \ref{eq:diff_gs_conn}, we
find 
\begin{equation}
\overline{A_{1}}+\left(\overline{A_{0}B_{1}}\right)_{d}=-it_\mathrm{r}\sum_{\ell}e^{-i\ell\Omega t_\mathrm{r}}p_{0\ell}\left[\partial_{\lambda}\varphi^{(\ell)}-\sum_{\ell^{\prime}}p_{0\ell^{\prime}}\partial_{\lambda}\varphi^{(\ell^{\prime})}\right].\label{eq:berry_conn_diff}
\end{equation}
In words the $\ell$-th peak is shifted by an amount proportional
to the difference between its Berry connection, $\partial_{\lambda}\varphi^{(\ell)}$,
and the mode-averaged Berry connection, $\sum_{\ell^{\prime}}p_{0}^{(\ell^{\prime})}\partial_{\lambda}\varphi^{(\ell)}$.
This is surprising, as the Berry connection is not gauge invariant
and thus observables expressed in terms of it seem not gauge invariant
on their face. However, the term above is in fact gauge invariant,
which comes from the fact that all of the Fourier modes are shifted
by the same the phase. To see this, consider a new gauge choice $|0_{F}^{\prime}(\lambda,t)\rangle=e^{i\chi(\lambda)}|0_{F}(\lambda,t)\rangle$.
Then 
\begin{equation}
|0_{F}^\prime(\lambda,t)\rangle=\sum_{\ell}e^{i\ell\Omega t}|0_{F}^{\prime(\ell)}\rangle=e^{i\chi(\lambda)}\sum_{\ell}e^{i\ell\Omega t}|0_{F}^{(\ell)}\rangle\implies|0_{F}^{\prime(\ell)}\rangle=e^{i\chi(\lambda)}|0_{F}^{(\ell)}\rangle~.
\end{equation}
But then $\varphi^{(\ell)}\to\varphi^{(\ell)}+\chi$ and the $\chi$
contribution will clearly drop out in Eq. \ref{eq:berry_conn_diff},
since $\sum_{\ell^{\prime}}p_{0\ell^{\prime}}=1$.

Meanwhile, the off-diagonal terms in $\overline{A_{0}B_{1}}$ can
be made to look more like $\overline{A_{2}}$. By first exchanging
the indices $\ell^{\prime}$ and $\ell^{\prime\prime}$ in the second term followed by
using the fact that $\tilde{A}_{F}^{(-\ell)}=\tilde{A}_{F}^{(\ell)\dagger}$
from the fact that $\tilde{A}_{F}(t)$ is Hermitian, we find that
\begin{eqnarray}
\left(\overline{A_{0}B_{1}}\right)_{od} & = & \sum_{\ell^{\prime}\neq\ell^{\prime\prime}}\frac{\langle e_{0}|\tilde{A}_{F}^{(\ell^{\prime}-\ell^{\prime\prime})}|e_{0}\rangle\langle e_{0}|V^{(\ell^{\prime})\dagger}V^{(\ell^{\prime\prime})}|e_{0}\rangle}{(\ell^{\prime}-\ell^{\prime\prime})\Omega}\left(e^{-i\ell^{\prime}\Omega t_\mathrm{r}}-e^{-i\ell^{\prime\prime}\Omega t_\mathrm{r}}\right)\\
 & = & \sum_{\ell^{\prime}\neq\ell^{\prime\prime}}e^{-i\ell^{\prime}\Omega t_\mathrm{r}}\left(\frac{\langle e_{0}|\tilde{A}_{F}^{(\ell^{\prime}-\ell^{\prime\prime})}|e_{0}\rangle\langle e_{0}|V^{(\ell^{\prime})\dagger}V^{(\ell^{\prime\prime})}|e_{0}\rangle}{(\ell^{\prime}-\ell^{\prime\prime})\Omega}+h.c.\right)\\
 & = & \sum_{\ell^{\prime}}e^{-i\ell^{\prime}\Omega t_\mathrm{r}}\sum_{\ell\neq0}\left(\frac{\langle e_{0}|\tilde{A}_{F}^{(\ell)}|e_{0}\rangle\langle e_{0}|V^{(\ell^{\prime})\dagger}V^{(\ell^{\prime}-\ell)}|e_{0}\rangle}{\ell\Omega}+h.c.\right).
\end{eqnarray}
Adding this to $\overline{A_{2}}$, we find that 
\begin{equation}
\overline{A_{2}}+\left(\overline{A_{0}B_{1}}\right)_{od}=\sum_{\ell^{\prime}}e^{-i\ell^{\prime}\Omega t_\mathrm{r}}\sum_{(n,\ell)\neq(0,0)}\left(\frac{\langle e_{0}|V^{(\ell^{\prime})\dagger}V^{(\ell^{\prime}-\ell)}|e_{n}\rangle\langle e_{n}|\tilde{A}_{F}^{(\ell)}|e_{0}\rangle}{\epsilon_{n0}^{F}+\ell\Omega}+h.c.\right).\label{eq:corr_off_diag}
\end{equation}

It bears mentioning that the frequency shift is zero at this order
in the undriven case. This can be seen from the above Floquet solution
by replacing the quasienergies $\epsilon_{n}^{F}$ with the actual
energies $E_{n}$ and only allowing $\ell,\ell^{\prime},\ell^{\prime\prime}=0$. Then the
Berry connection term (Eq. \ref{eq:berry_conn_diff}) vanishes because
one subtracts the Berry connection of the ground state from itself.
Similarly, the off-diagonal corrections (Eq. \ref{eq:corr_off_diag})
vanish because the term $\langle e_{0}|V^{(0)\dagger}V^{(0)}|e_{n}\rangle=\langle E_{0}|E_{n}\rangle=0$
from orthogonality of the energy eigenstates.

Finally, having solved for the Wigner distribution in terms of the average and relative times, we must Fourier transform and convolve with the probe to get the actual ARPES signal and see that there are no additional corrections to order $\dot \lambda$. We rewrite the diagonal (\eref{eq:berry_conn_diff}) and off-diagonal (\eref{eq:corr_off_diag}) corrections as $a_d$ and $a_{od}$ respectively, such that
\begin{equation}
\overline{f}(t_r,t_{av}) \approx e^{i \epsilon^F_0 t_r} \sum_\ell e^{-i \ell \Omega t_r} p_0^{(\ell)} \left[ 1 + \dot \lambda_{av} (a_{od}^{(\ell)} - i t_r a_d^{(\ell)}) \right].
\end{equation}
This trivially Fourier transformed to get
\begin{equation}
\overline{f}(\omega,t_{av}) \approx 2\pi \sum_\ell p_0^{(\ell)} \left[ (1 + \dot \lambda_{av} a_{od}^{(\ell)})\delta(\omega - \epsilon_0^F + \ell \Omega) + \dot \lambda_{av} a_d^{(\ell)} \delta'(\omega - \epsilon_0^F + \ell \Omega) \right].
\end{equation}
Now let us convolve this Wigner distribution by a Gaussian probe
to get the ARPES signal and confirm that these results are unaffected
by smearing the $\delta$-function peaks by Gaussians. First convolving
along the $\omega$ direction (see Eq. \ref{eq:def_I_omega_tav}),
we get 
\begin{eqnarray*}
I_{1}(\omega,t_\mathrm{av}) & \equiv & \int_{-\infty}^{\infty}d\omega^{\prime}\overline{f}(\omega^{\prime},t_\mathrm{av})e^{-(\omega^{\prime}-\omega)^{2}\tau_\mathrm{pr}^{2}}\\
 & \approx & 2\pi\sum_{\ell}p_{0\ell}\big[(1+\dot{\lambda}_\mathrm{av}a_{od}^{(\ell)})e^{-(\omega-\epsilon_{0}^{F}+\ell\Omega)^{2}\tau_\mathrm{pr}^{2}}\\
 &  & +2\dot{\lambda}_\mathrm{av}a_{d}^{(\ell)}(\omega-\epsilon_{0}^{F}+\ell\Omega)\tau_\mathrm{pr}^{2}e^{-(\omega-\epsilon_{0}^{F}+\ell\Omega)^{2}\tau_\mathrm{pr}^{2}}\big]\\
 & \approx & 2\pi\sum_{\ell}p_{0\ell}\left[(1+\dot{\lambda}_\mathrm{av}a_{od}^{(\ell)})e^{-(\omega-\epsilon_{0}^{F}+\ell\Omega-\dot{\lambda}_\mathrm{av}a_{d}^{(\ell)})^{2}\tau_\mathrm{pr}^{2}}\right],
\end{eqnarray*}
corresponding to a frequency shift of $\dot{\lambda}_\mathrm{av}a_{d}^{(\ell)}$.
Second, we must convolve in the time direction with the probe envelope
$e^{-(t_\mathrm{av}-t_\mathrm{pr})^{2}/\tau_\mathrm{pr}^{2}}$. The previous expression
for $I_{1}(\omega,t_\mathrm{av})$ only depends on $t_\mathrm{av}$ through $\lambda_\mathrm{av}$.
Therefore assuming that that probe is short such that $\lambda$ does
not significantly change during it (i.e., $\tau_{\mathrm{ramp}}\gg\tau_\mathrm{pr}$)
we must ask when it is appropriate to simply replace $t_\mathrm{av}$ by
$t_\mathrm{pr}$. This clearly correct for all terms of order $\dot{\lambda}$,
because doing a Taylor series in the difference $t_\mathrm{av}-t_\mathrm{pr}$ times
the derivative of these terms with respect to $\lambda$ would lead to corrections
of order $\dot{\lambda}^{2}$. Thus the only potentially relevant
correction comes from the term $p_{0\ell}(\lambda_\mathrm{av})\exp[-(\omega-\epsilon_{0}^{F}(\lambda_\mathrm{av})+\ell\Omega)]\equiv C_{0}(\lambda_\mathrm{av})$.
Fortunately, a Taylor expansion in $t_\mathrm{av}-t_\mathrm{pr}$ gives $\dot{\lambda}(t_\mathrm{pr})(t_\mathrm{av}-t_\mathrm{pr})C_{0}^{\prime}(\lambda_\mathrm{pr})$,
which is odd w.r.t. $(t_\mathrm{av}-t_\mathrm{pr})$ and thus vanishes under integration
with the Gaussian. So at order $\dot{\lambda}$ we get our final answer
for the ARPES signal:
\begin{equation}
I(\omega,t_\mathrm{pr})\approx\sum_{\ell}p_{0\ell}\left[(1+\dot{\lambda}_\mathrm{pr}a_{od}^{(\ell)}(\lambda_\mathrm{pr}))e^{-[\omega-\epsilon_{0}^{F}(\lambda_\mathrm{pr})+\ell\Omega-\dot{\lambda}_\mathrm{pr}a_{d}^{(\ell)}(\lambda_\mathrm{pr})]^{2}\tau_\mathrm{pr}^{2}}\right]~,\label{eq:S_omega_tpr}
\end{equation}
which is the final result reproduced in Eq.~\ref{eq:I_om_tpr_f}.

\end{document}